\documentclass[review,3p,times]{elsarticle}
\makeatletter
\def\@blstr{1.2}
\makeatother
\usepackage{setspace}
\usepackage[T1]{fontenc}
\usepackage[utf8]{inputenc}
\usepackage{amsmath,amssymb}
\usepackage{newtxmath}
\usepackage{microtype}
\usepackage{booktabs,tabularx,longtable,array}
\usepackage{graphicx,float,caption,xcolor,enumitem}
\usepackage[hidelinks]{hyperref}
\biboptions{numbers,sort&compress}
\definecolor{navy}{HTML}{16324F}
\definecolor{pale}{HTML}{EEF3F8}
\hypersetup{colorlinks=true,linkcolor=navy,citecolor=navy,urlcolor=navy,
  pdftitle={Which Past Matters? Training History and Multi-Forecast Horizon in Equity Transformers},
  pdfauthor={Chariton Chalvatzis},
  pdfsubject={Historical sampling and intraday equity forecasting},
  pdfkeywords={Equity forecasting, Transformers, Historical sampling, Forecast horizon, Concept drift}
}
\setlist{nosep}
\setlist[itemize,enumerate]{leftmargin=*}
\floatstyle{plain}
\restylefloat{table}
\restylefloat{figure}
\usepackage{xurl}
\begin{document}
\begin{frontmatter}
\title{Which Past Matters? Training History and Multi-Forecast Horizon in Equity Transformers}
\author[uom]{Chariton Chalvatzis}
\ead{charis.chalvatzis@mfe.berkeley.edu}
\address[uom]{University of Macedonia, Department of Applied Informatics, 156 Egnatia St., Thessaloniki, 54006, Greece}

\begin{abstract}
More historical data can improve estimation while introducing relationships that transfer poorly to current market conditions. We therefore examine how training quantity and historical selection interact with forecast horizon in intraday equity prediction. We deploy twelve configurations which combine four sampling and validation policies with 6-, 12- and 24-month training quantities under a fixed cross-sectional Transformer and feature set. A monthly walk-forward experiment produces 264 refits and eight cumulative return forecasts from 5 to 120 minutes. Evaluation covers January 2025 through 20 October 2026 and uses cross-sectional Spearman information coefficients against future five-factor residual returns. Six-month recency sampling leads individual models at 5--15 minutes, while multi-recency configurations lead at 20--30 minutes and uniform configurations at 60--120 minutes. Increasing training quantity does not uniformly improve accuracy while an equal-weight ensemble of all twelve models exceeds every individual mean at 15, 20 and 25 minutes, but not at the remaining horizons. These findings support evaluating historical selection jointly with forecast horizon without identifying a universal policy or a causal benefit from recency. 
\end{abstract}
\end{frontmatter}

\section{Introduction}
Increasing the amount of training data is a central strategy for improving machine-learning models. In language modelling, empirical scaling laws document systematic performance improvements as training data, model capacity and computation increase \cite{kaplan2020}. These findings motivate the pursuit of larger datasets but do not establish that every additional observation is equally useful for every prediction task. In financial forecasting, expanding a training sample generally also expands its historical coverage, potentially introducing observations generated under conditions different from those prevailing at the forecast date.

Time introduces two distinct considerations. First, a forecasting experiment must respect the information available when a prediction would have been issued: training observations, labels and preprocessing estimates must not incorporate information from the subsequent evaluation period. Second, even a strictly chronological experiment faces a question of historical relevance. Observations that were legitimately available may nevertheless describe relationships that have weakened, changed or reversed. Preventing future-information leakage therefore establishes the validity of the information set but does not determine which parts of that information set are most useful.

This creates a trade-off between the quantity and relevance of training observations. A longer history can improve estimation and broaden the range of market conditions represented in training. However, it can also introduce relationships that transfer poorly to the current prediction period. Conversely, restricting training to recent observations may improve alignment with current conditions while increasing estimation uncertainty and reducing exposure to less frequently observed states. Older observations should therefore neither be retained automatically nor discarded solely because of their age. Research on structural change demonstrates that observations preceding a break can remain useful when their contribution to reducing estimation variance outweighs the bias they introduce \cite{pesaran2007}.

The balance between these considerations may depend on the forecast horizon. Historical age measures how far a training observation lies in the past while forecast horizon measures how far the target lies in the future. These are distinct dimensions and their interaction is an empirical question since relationships useful for predicting returns over a few minutes may evolve differently from those relevant over longer intervals. Consequently, a training sample appropriate for one horizon need not be equally effective at another. This motivates investigating horizon-dependent historical relevance rather than assigning a universal threshold beyond which observations become \textquotedblleft old\textquotedblright.

Training-data selection is therefore part of the forecasting model's design. A recent contiguous window emphasizes the latest observations, uniform historical sampling broadens coverage and recency-weighted sampling balances a preference for newer observations with access to older periods. These policies construct different training distributions, even when the model architecture and nominal sample size are unchanged. Their purpose is to investigate the balance between relevance and historical diversity, rather than to assume that either recency or diversity necessarily improves prediction.

We therefore ask: \textbf{how do training quantity and historical sampling policy interact with forecast horizon in intraday equity forecasting?} We investigate this question using a fixed cross-sectional Transformer and feature set. The model attends across contemporaneous stocks while historical information enters through engineered features and the observations selected for training. The study thus examines the role of training history without requiring temporal attention or an explicit regime-identification mechanism.

Our methodology uses a monthly walk-forward design issuing forecasts at 30-minute intervals for up to 1,000 liquid US equities represented by 572 engineered features. Four sampling and validation policies combine with 6-, 12- and 24-month training quantities to give twelve configurations. These policies use contiguous recent history, uniform month sampling, recency-weighted month sampling or multi-recency sampling with several chronological validation blocks. The corresponding historical pools span 12, 24 and 36 months. Earlier validation periods determine the number of training epochs before a fresh model is fitted on the selected history together with the latest validation block and frozen for the next test month. Repeating this procedure over 22 test months produces 264 final refits. Each forecast origin has eight cumulative return horizons ($5/10/15/20/25/30/60/120$ minutes), excluding targets that extend beyond the session close.

The primary evidence is the cross-sectional Spearman rank information coefficient (Rank IC) against future five-factor residual returns. IC is computed within each forecast origin, averaged within each session and then averaged across sessions. The main comparisons cover January 2025 through 20 October 2026: 451 sessions, 5,394 forecast origins and 5,248,638 stock--origin rows per configuration. Outcomes enter evaluation subject to the market-close mask and exact-endpoint label availability. An ensemble averages all twelve models' predictions with equal weights for each stock, forecast origin and horizon.

The results show that the leading individual configuration depends on forecast horizon. Six-month recency sampling leads at 5--15 minutes with mean Rank IC of 1.90\%, 1.57\% and 1.00\%, respectively. Multi-recency configurations lead at 20--30 minutes while uniform configurations lead at 60 and 120 minutes. Increasing training quantity does not uniformly improve accuracy. Within recency sampling, the leading quantity progresses from 6 months at 5--15 minutes to 12 months at 20--30 minutes and 24 months at 60--120 minutes but this pattern does not hold across all policy families.

The ensemble exceeds every individual mean at 15, 20 and 25 minutes but not at the other horizons. Performance also varies substantially across calendar periods: ensemble 5-minute Rank IC falls from 3.72\% in H1 2025 to 0.28\% in H2 before reaching 1.01\% in 2026 through 20 October. These results distinguish full-period predictive rankings from stable superiority over time.

Our contribution is a common-model empirical framework for evaluating historical-data selection jointly with intraday forecast horizon. By comparing four explicit sampling and validation policies at three training quantities across eight horizons, we examine the consequences of changing training history while holding the architecture, features and forecast cadence fixed. The comparison connects individual-model accuracy with forecast combination and calendar-period stability. It extends estimation-window research under structural change \cite{pesaran2007,inoue2017} to nonlinear cross-sectional intraday forecasting and complements learned data-adaptation methods \cite{li2022,zhao2023doubleadapt,taga2026faf} through transparent historical-selection rules. The resulting evidence identifies horizon-dependent choices worth testing prospectively. It does not isolate a causal effect of recency because historical pool breadth changes with training quantity and validation procedures differ across policy families.

The remainder of this study is organized as follows. Section~\ref{sec:literature} reviews the related literature. Section~\ref{sec:experimental-design} presents the experimental design and Section~\ref{sec:architecture} describes the Transformer architecture and training procedure. Section~\ref{sec:results} defines the evaluation methodology and presents the results while Section~\ref{sec:conclusions} draws the conclusions and discusses potential improvements.

\section{Literature Review}
\label{sec:literature}
The choice of training history connects several strands of research. Studies of forecasting under structural change examine the trade-off between additional observations and their relevance, while adaptive time-series methods seek to adjust training data or models as distributions evolve. Work on language-model data selection offers a complementary perspective on how relevance, mixture composition and redundancy affect the value of a dataset. We review these approaches to position our comparison of historical sampling policies and to distinguish their objectives from our question of how useful training history varies with intraday forecast horizon.

The study in \cite{pesaran2007} examines training quantity under structural change. It shows that including observations from before a break can improve forecasts despite introducing bias, because the additional data can reduce estimation variance. The authors develop cross-validation and forecast-pooling procedures for uncertain break dates and magnitudes. This provides a theoretical motivation for comparing recent-only history with samples that retain older observations. Our R6--C6 comparison addresses a related trade-off, although it does not isolate the same mechanism. We sample separated months, use a nonlinear model and evaluate cross-sectional IC, whereas the study's principal criterion is mean-squared forecast error.

Extending this question from sample quantity to observation weights, the authors in \cite{pesaran2013} derive weights for one-step-ahead prediction under continuous and discrete changes. Continuous change largely recovers exponential smoothing, while discrete breaks imply weights that differ across regimes. Robust procedures are also proposed when break information is uncertain. This motivates our recency preference while showing why calendar age need not fully describe relevance. Our procedure uses exponential weights to select whole months without replacement and then trains without additional age weights, so it differs from the weighted estimator considered in that study.

A related approach in \cite{inoue2017} treats the estimation window as a data-dependent choice. The authors develop a rolling-window selection methodology that minimizes quadratic forecast loss, with asymptotic justification and macroeconomic applications. Its predictive-regression framework allows horizon-specific relationships, making it relevant to whether useful training history changes with forecast horizon. It does not imply that longer horizons necessarily require more history. Our discrete policy grid examines this question through nonlinear cross-sectional forecasts and a different evaluation criterion.

Rather than selecting a single history, the study in \cite{pesaran2011} combines forecasts from the same model fitted over different estimation windows. Theoretical analysis under structural breaks and an application to weekly equity-index-futures returns support combining windows to reduce forecast error in the settings examined. This is directly relevant to our equal ensemble, because historical estimation samples can generate forecast diversity even when architecture and features are fixed. Our ensemble additionally combines sampling and validation policies, making its horizon-dependent performance an empirical question.

Adaptive forecasting methods extend these ideas by learning how historical observations should contribute to training. The authors in \cite{li2022} estimate evolving data distributions and construct training samples through historical resampling, with experiments that include stock-price-trend forecasting. Their approach motivates learning historical relevance instead of imposing only uniform or age-based selection. Such a sampling rule would require chronological validation and an untouched evaluation period in our setting.

The study in \cite{zhao2023doubleadapt} extends adaptation to both the data and the forecasting model. It learns adapters that mitigate distribution shifts during incremental updates, providing a closely related application in stock forecasting. Our experiment instead refits a fixed architecture on explicitly selected historical months, without learning data transformations or parameter-update rules. This distinction suggests an adaptive benchmark for future comparisons rather than evidence about which history is best at our particular intraday horizons.

Training-data construction is also central to \cite{taga2026faf}, where reference-model-based filtering is combined with augmentation of retained examples. The proposed adaptive method updates its reference model to address limitations of holdout-loss-based selection under distribution shifts. This offers another way to improve forecasting through the training sample while retaining the forecasting architecture. Its emphasis on example quality and augmentation differs from our focus on historical coverage and calendar-age preferences and does not imply that older observations are intrinsically noisy.

A complementary literature examines data selection for language models. The study in \cite{xie2023dsir} estimates importance weights in a reduced feature space and resamples text to approximate a target distribution specified by unlabeled examples. This provides a methodological parallel to selecting historical observations for relevance. An extension to financial forecasting could estimate relevance from market-state features rather than calendar age alone, provided that the target-distribution estimate uses only information available at the forecast date.

Moving from individual examples to the mixture of training domains, the authors in \cite{xie2023doremi} use a small proxy model and group distributionally robust optimization to estimate domain proportions before resampling data for a larger language model. Their findings show that the allocation of training exposure affects performance and efficiency. The corresponding question in our setting is how to allocate exposure across historical months or market states. These temporal groups are not equivalent to text domains, so the approach motivates an extension without establishing an optimal recency rule.

The value of additional data also depends on redundancy. The study in \cite{tirumala2023d4} uses pretrained embeddings to combine semantic deduplication with diversified document selection, improving language-model training efficiency and downstream performance in its experiments. This motivates examining whether a larger financial sample adds distinct information or mainly repeats similar conditions. However, serially correlated financial observations are not equivalent to duplicate documents, because removing them changes the empirical frequency of market states.

Together, these studies distinguish observation weighting, sample selection, validation design and forecast combination as separate modelling choices. They motivate our comparison of explicit historical-selection rules across intraday horizons while providing alternatives for future adaptive benchmarks.

\section{Experimental design}
\label{sec:experimental-design}

We examine whether the useful age and quantity of training data depend on the horizon of an equity-return forecast. A fixed cross-sectional Transformer is refitted monthly under twelve historical-selection configurations. The configurations vary training quantity (6, 12 or 24 months), historical sampling (contiguous, uniform or recency weighted) and, for one policy family, the number of chronological validation blocks. We first define the data, universe and forecast cadence, then distinguish selected training history from its eligible historical pool. We next specify the sampling and validation policies and illustrate their implementation for a January 2025 forecast.

\subsection{Data and prediction cadence}
\textbf{Data source.} Market data come from Massive (formerly Polygon), using its US-equity SIP one-minute and daily aggregate flat files. Raw OHLCV bars are retained unadjusted. The research pipeline constructs adjusted prices using split and cash-dividend reference data. Minute bars supply intraday features and return endpoints while daily bars support liquidity screening and the five-factor risk model. The 572 inputs are bar-derived features, not trade-by-trade order flow or quoted order-book measures.

\textbf{Universe.} At the first trading session of each month, select up to 1,000 common stocks ranked by trailing 90-trading-session average dollar volume (ADV). Require a prior-session unadjusted close of at least \$5, ADV of at least \$10 million and valid positive dollar volume in all 90 prior sessions. Daily dollar volume is approximated by unadjusted close times share volume. Membership is held fixed within the month and selection uses information available before the rebalance session. The reference filter uses instrument type \texttt{CS} across active and inactive ticker lists, excluding ETFs and other security types. These price/liquidity rules are lagged, but the reference classification itself is not independently certified here as a historical security master. Actual model-row availability remains a separate restriction.

\textbf{Feature groups.} The library contains 72 base inputs and 500 deterministic extensions. The base set comprises return, volatility, volume and liquidity context (14); market, momentum, reversal, volatility and liquidity risk exposures (5); total-return and residual mean reversion (26); broad, correlation, ridge and liquidity-basket relative value (12); acceleration, VWAP displacement and volume climax (4); intraday/weekday calendar encodings (4); and seven alpha-cluster composites. Extensions reuse 20 anchor signals, so 572 columns do not represent 572 independent economic hypotheses.
Table~\ref{tab:feature-groups} summarizes these feature groups.
\begin{table}[htbp]\centering\small
\caption{Feature groups in the 572-input library. Extensions are deterministic transformations of shared anchor signals.}
\label{tab:feature-groups}
\renewcommand{\arraystretch}{1.1}
\begin{tabular}{lrp{0.59\linewidth}}\toprule
Group & Count & Examples\\\midrule
Base inputs & 72 & Reversal, basket dislocations, risk exposures and context\\
Temporal transforms & 160 & Changes, moving means, history z-scores and acceleration\\
Nonlinear transforms & 80 & Squares, cubes, signed squares and tail excess\\
Context interactions & 160 & Products or damping by volatility, volume, liquidity and styles\\
Conditional signals & 40 & Reversal gated by volatility or liquidity regimes\\
Peer-relative signals & 48 & Deviations from liquidity, volatility, momentum or fitted-cluster peers\\
Completed-return signals & 12 & Observed residual reversal, shocks, skew and same-clock-time history\\\bottomrule
\end{tabular}
\end{table}
The experiment applies the 572-feature library to the monthly universe defined above. Actual model rows depend on feature availability; there are approximately 973 stock rows per forecast origin on average. The input preparation checks that any frozen monthly cluster fit ends before the test month. 

Forecasts are issued at 30-minute intervals, beginning at 10:00 New York time on regular trading days. We exclude the first 30 minutes after the market opens to allow the initial price-discovery process to unfold. They predict cumulative residual returns over
\[
\mathcal H=\{5,10,15,20,25,30,60,120\}\text{ minutes}.
\]
A 5-minute output is not a forecast issued every five minutes: the stored origin cadence remains 30 minutes. A horizon extending beyond the session close is masked as time-ineligible. Targets do not cross overnight. The 60-minute target covers $[t,t+60]$, not just the increment after $t+30$ and the 120-minute target likewise starts at $t$.

\subsection{Training quantity and historical pool}
For each monthly forecasting exercise, we distinguish the amount of history used to fit the model from the span of history eligible for selection. The historical pool consists of $H$ consecutive calendar months ending immediately before a two-month validation block. These pool months are candidate training periods, not additional observations automatically included in training. A policy selects $T$ distinct months from this pool, and the available observations within those selected months form its selection-training sample while unselected months are excluded from that fit. Thus, $T$ controls the quantity of selected history, whereas $H$ controls how far back the selection can reach. For example, $T=6$ and $H=12$ means training on six months chosen from a twelve-month window; the selected months need not be consecutive. Contiguous selection uses only the newest $T$ months, while sampled policies can include older months from the same eligible pool. Table~\ref{tab:training-families} summarizes the three training-history families. The short family selects six months from a twelve-month pool, the medium family selects twelve from twenty-four months, and the long family selects twenty-four from thirty-six months. These names describe the quantity of selected training history, not the forecast horizon. They provide progressively larger samples and broader historical coverage; because $T$ and $H$ change together, their comparison does not isolate training quantity alone.
\begin{table}[htbp]\centering
\caption{Training-history families and eligible historical pools. Final refits add the latest two validation months to the selected training sample.}
\label{tab:training-families}
\renewcommand{\arraystretch}{1.15}
\begin{tabular}{lrrr}\toprule
Family & Selected training months $T$ & Pool months $H$ & Final refit months\\\midrule
Short & 6 & 12 & 8\\
Medium & 12 & 24 & 14\\
Long & 24 & 36 & 26\\\bottomrule
\end{tabular}\end{table}

The validation block is held out during selection training and determines the number of training epochs. After this choice, a newly initialized model is fitted on the selected $T$ months together with the latest two validation months, giving $T+2$ months in the final refit. This model is then frozen for the following test month. Consequently, the six-month family uses six selected training months but eight months in its final fit. Validation observations enter the final refit only after their role in epoch selection is complete; the test month remains excluded throughout.

\subsection{Four sampling and validation policies}
Within each training-quantity family, we compare four ways of constructing and validating the historical sample. The first three share a single two-month validation block and differ in which months they select: consecutive recent history, an age-neutral sample, or a sample favouring recent history. The fourth retains recency sampling but uses several chronological validation blocks to make epoch selection less dependent on one period. These rules separate the selection of historical observations from the subsequent fitting of the same forecasting architecture.

\begin{description}[style=nextline,leftmargin=1.5em,itemsep=0.65\baselineskip,parsep=0.2\baselineskip,topsep=0.5\baselineskip]
\item[Contiguous (C).] Select the most recent $T$ consecutive months before validation. The older part of the nominal pool is unused.
\item[Uniform (U).] Select $T$ distinct months without replacement from the $H$-month pool, with equal selection probability.
\item[Recency (R).] Select $T$ distinct months without replacement, with month-selection weights
\[
q_a\propto 2^{-a/6},\qquad a=0,\ldots,H-1,
\]
where $a=0$ is the newest pool month. This is a six-month half-life for sampling probabilities, not an exponential observation-weighted training loss. Once a month is selected, its available snapshots enter training without an additional recency weight.
\item[Multi-recency (M).] Repeat recency sampling across three chronological two-month validation blocks. Fit each selection model independently with early stopping. Take the median of their best epochs, then freshly refit once on the latest block's selected training months plus its validation months. This produces one final model, not an average of three fitted models.
\end{description}

Model labels combine the policy letter and $T$: R6 is six-month recency, M24 is 24-month multi-recency. There are $3\times4=12$ variants per test month, and $12\times22=264$ final refits. The model initialization seed is 7 throughout. Sampling uses deterministic job-specific draws derived from sampling seed 101; epoch-level snapshot shuffling uses seed $1001+e$. Architecture, features, label definition, initialization seed, optimizer hyperparameters and forecast cadence are held fixed. Comparing C6/C12/C24 isolates different contiguous history lengths more cleanly than comparing sampled quantities. For U, R and M, increasing $T$ also changes $H$ from 12 to 24 to 36 months. Training quantity, historical breadth and sampling fraction therefore move together. M also differs in validation design. These are structured policy comparisons and do not constitute a fully crossed causal experiment.

\subsubsection{Concrete January 2025 example}
\begin{figure}[htbp]
\centering
\setlength{\unitlength}{1mm}
\resizebox{\linewidth}{!}{%
\begin{picture}(184,82)
\scriptsize
\put(40,70){\makebox(108,8){Historical pool: 12 months}}
\put(40,68){\line(1,0){108}}
\put(40,65){\line(0,1){3}}
\put(148,65){\line(0,1){3}}
\put(149,69){\makebox(17,9){\shortstack{Validation\\2 months}}}
\put(149,68){\line(1,0){17}}
\put(167,69){\makebox(17,9){\shortstack{Test\\Jan 2025}}}
\put(167,68){\line(1,0){9}}
\put(0,53){\makebox(38,6)[r]{C6: latest six months}}
\put(0,43){\makebox(38,6)[r]{R6: illustrative draw}}
\put(0,7){\makebox(38,6)[r]{C6 final refit}}
\put(0,-3){\makebox(38,6)[r]{R6 final refit}}
\put(40,30){\line(0,1){2}}
\put(40,24){\makebox(9,5){Nov}}
\put(40,53){\color{black!20}\framebox(9,6){}}
\put(40,43){\color{black!20}\framebox(9,6){}}
\put(49,30){\line(0,1){2}}
\put(49,24){\makebox(9,5){Dec}}
\put(49,53){\color{black!20}\framebox(9,6){}}
\put(49,43){\color{black!20}\framebox(9,6){}}
\put(58,30){\line(0,1){2}}
\put(58,24){\makebox(9,5){Jan}}
\put(58,53){\color{black!20}\framebox(9,6){}}
\put(58,43){\color{black!20}\framebox(9,6){}}
\put(67,30){\line(0,1){2}}
\put(67,24){\makebox(9,5){Feb}}
\put(67,53){\color{black!20}\framebox(9,6){}}
\put(67,43){\color{black!20}\framebox(9,6){}}
\put(76,30){\line(0,1){2}}
\put(76,24){\makebox(9,5){Mar}}
\put(76,53){\color{black!20}\framebox(9,6){}}
\put(76,43){\color{black!20}\framebox(9,6){}}
\put(85,30){\line(0,1){2}}
\put(85,24){\makebox(9,5){Apr}}
\put(85,53){\color{black!20}\framebox(9,6){}}
\put(85,43){\color{black!20}\framebox(9,6){}}
\put(94,30){\line(0,1){2}}
\put(94,24){\makebox(9,5){May}}
\put(94,53){\color{black!20}\framebox(9,6){}}
\put(94,43){\color{black!20}\framebox(9,6){}}
\put(103,30){\line(0,1){2}}
\put(103,24){\makebox(9,5){Jun}}
\put(103,53){\color{black!20}\framebox(9,6){}}
\put(103,43){\color{black!20}\framebox(9,6){}}
\put(112,30){\line(0,1){2}}
\put(112,24){\makebox(9,5){Jul}}
\put(112,53){\color{black!20}\framebox(9,6){}}
\put(112,43){\color{black!20}\framebox(9,6){}}
\put(121,30){\line(0,1){2}}
\put(121,24){\makebox(9,5){Aug}}
\put(121,53){\color{black!20}\framebox(9,6){}}
\put(121,43){\color{black!20}\framebox(9,6){}}
\put(130,30){\line(0,1){2}}
\put(130,24){\makebox(9,5){Sep}}
\put(130,53){\color{black!20}\framebox(9,6){}}
\put(130,43){\color{black!20}\framebox(9,6){}}
\put(139,30){\line(0,1){2}}
\put(139,24){\makebox(9,5){Oct}}
\put(139,53){\color{black!20}\framebox(9,6){}}
\put(139,43){\color{black!20}\framebox(9,6){}}
\put(148,30){\line(0,1){2}}
\put(148,24){\makebox(9,5){Nov}}
\put(157,30){\line(0,1){2}}
\put(157,24){\makebox(9,5){Dec}}
\put(166,30){\line(0,1){2}}
\put(166,24){\makebox(9,5){Jan}}
\put(40,30){\vector(1,0){139}}
\put(40,19){\makebox(18,4){2023}}
\put(58,19){\makebox(108,4){2024}}
\put(166,19){\makebox(9,4){2025}}
\put(40,43){\color{orange!80!black}\rule{9\unitlength}{6\unitlength}}
\put(40,-3){\color{orange!80!black}\rule{9\unitlength}{6\unitlength}}
\put(67,43){\color{orange!80!black}\rule{9\unitlength}{6\unitlength}}
\put(67,-3){\color{orange!80!black}\rule{9\unitlength}{6\unitlength}}
\put(94,53){\color{blue!65!black}\rule{9\unitlength}{6\unitlength}}
\put(94,43){\color{orange!80!black}\rule{9\unitlength}{6\unitlength}}
\put(94,7){\color{blue!65!black}\rule{9\unitlength}{6\unitlength}}
\put(94,-3){\color{orange!80!black}\rule{9\unitlength}{6\unitlength}}
\put(103,53){\color{blue!65!black}\rule{9\unitlength}{6\unitlength}}
\put(103,7){\color{blue!65!black}\rule{9\unitlength}{6\unitlength}}
\put(112,53){\color{blue!65!black}\rule{9\unitlength}{6\unitlength}}
\put(112,43){\color{orange!80!black}\rule{9\unitlength}{6\unitlength}}
\put(112,7){\color{blue!65!black}\rule{9\unitlength}{6\unitlength}}
\put(112,-3){\color{orange!80!black}\rule{9\unitlength}{6\unitlength}}
\put(121,53){\color{blue!65!black}\rule{9\unitlength}{6\unitlength}}
\put(121,7){\color{blue!65!black}\rule{9\unitlength}{6\unitlength}}
\put(130,53){\color{blue!65!black}\rule{9\unitlength}{6\unitlength}}
\put(130,43){\color{orange!80!black}\rule{9\unitlength}{6\unitlength}}
\put(130,7){\color{blue!65!black}\rule{9\unitlength}{6\unitlength}}
\put(130,-3){\color{orange!80!black}\rule{9\unitlength}{6\unitlength}}
\put(139,53){\color{blue!65!black}\rule{9\unitlength}{6\unitlength}}
\put(139,43){\color{orange!80!black}\rule{9\unitlength}{6\unitlength}}
\put(139,7){\color{blue!65!black}\rule{9\unitlength}{6\unitlength}}
\put(139,-3){\color{orange!80!black}\rule{9\unitlength}{6\unitlength}}
\put(148,7){\color{black!45}\rule{9\unitlength}{6\unitlength}}
\put(148,-3){\color{black!45}\rule{9\unitlength}{6\unitlength}}
\put(157,7){\color{black!45}\rule{9\unitlength}{6\unitlength}}
\put(157,-3){\color{black!45}\rule{9\unitlength}{6\unitlength}}
\put(148,40){\color{black!45}\rule{18\unitlength}{21\unitlength}}
\put(148,47){\makebox(18,8){\shortstack{Select\\epochs}}}
\put(166,40){\framebox(9,21){\shortstack{Frozen\\model}}}
\end{picture}}
\caption{Monthly timeline for January 2025 forecasting. Above the calendar, C6 selects the newest six consecutive pool months (blue), whereas R6 selects six distinct months from the same pool (orange; an illustrative draw, not the recorded experimental draw). Both validate on November--December 2024 (grey). Below the calendar, each final-refit row identifies its six selected months plus the two validation months: these bars denote data membership, not when fitting occurs. A fresh model is refitted after validation and frozen before January test-month forecasts.}
\label{fig:january-selection}
\end{figure}
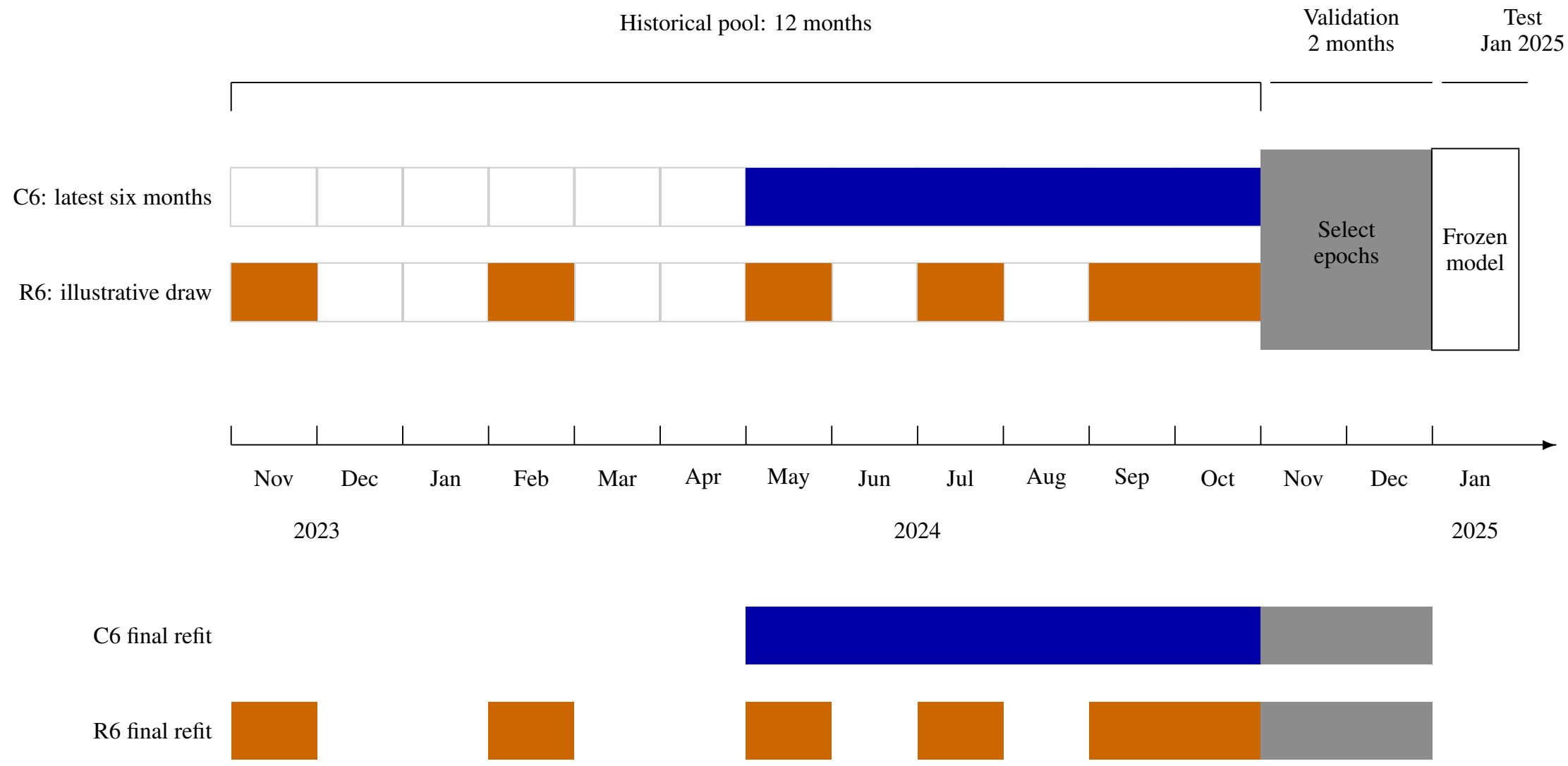

The chronological order for January 2025 forecast is given in Figure~\ref{fig:january-selection}. For example in the case of C6, we train on data from May to October 2024, validate on data from November to December 2024, then refit a newly initialized model on data from May to December 2024 for the selected number of epochs. Freeze it for January 2025 predictions. In case of R6, we draw six months using data from November 2023 to October 2024, validate on data from November to December 2024, then refit on those selected months plus validation.
M6 uses validation blocks from July to August, September to October and November to December 2024. Each has its own preceding pool and independent selection fit. The median of the three best epoch counts determines the duration of the final refit. Only the latest block's selected training sample and its November--December validation months enter that refit while earlier blocks contribute to epoch selection, not additional final-fit observations. Test months are never used in selection or refit.

\subsection{Return labels and residualization}
\label{sec:return-labels}
The forecasting targets are horizon-specific residual returns. Their construction removes common risk-factor movements while their evaluation must account for incomplete realized outcomes. For each origin and horizon, realized returns are cross-sectionally residualized using a five-factor weighted least squares (WLS) routine and lagged daily exposures. Its factor structure consists of a market/intercept exposure plus momentum, short-term reversal, volatility and liquidity styles. In schematic form,
\begin{align}
\widehat f_{t,h}&=\arg\min_f\sum_{i\in\mathcal R_t\cap\mathcal O_{t,h}}
\omega_{i,t}\bigl(r_{i,t,h}-b_{i,t}^{\top}f\bigr)^2,\\
y_{i,t,h}&=r_{i,t,h}-b_{i,t}^{\top}\widehat f_{t,h}.
\end{align}
Here $r_{i,t,h}$ is the adjusted total return for stock $i$ from forecast origin $t$ over horizon $h$, $b_{i,t}$ is known before the return period, $\mathcal R_t$ is the lagged risk universe, $\mathcal O_{t,h}$ is the set of stocks with observed returns for that origin and horizon, and $\omega_{i,t}$ is the normalized WLS regression weight. Let $\operatorname{ADV}_{i,90}$ be the trailing 90-trading-session average dollar volume recorded in the monthly universe-selection snapshot. Define
\begin{equation}
a_{i,t}=\min\!\left\{\sqrt{\operatorname{ADV}_{i,90}},\ Q_{0.99,t}\!\left(\sqrt{\operatorname{ADV}_{90}}\right)\right\},
\qquad \omega_{i,t}=\frac{a_{i,t}}{\sum_j a_{j,t}},
\end{equation}
where $Q_{0.99,t}$ is the cross-sectional 99th percentile of square-root ADV in that snapshot. The weights favour more liquid stocks while capping the influence of the largest values. Before capping, a stock with \$400 million ADV receives twice the weight of one with \$100 million ADV. These weights enter the residualization regression and do not define portfolio allocations.

 Realized factor returns are part of the label construction and not input features for prediction. Residualization is performed separately for each horizon; therefore, a 60-minute residual is not necessarily the sum or compounded product of two independently estimated 30-minute residuals. The outputs are horizon-specific residual targets; this study does not train disjoint-return heads.

\begin{figure}[htbp]\centering
\includegraphics[width=0.72\textwidth]{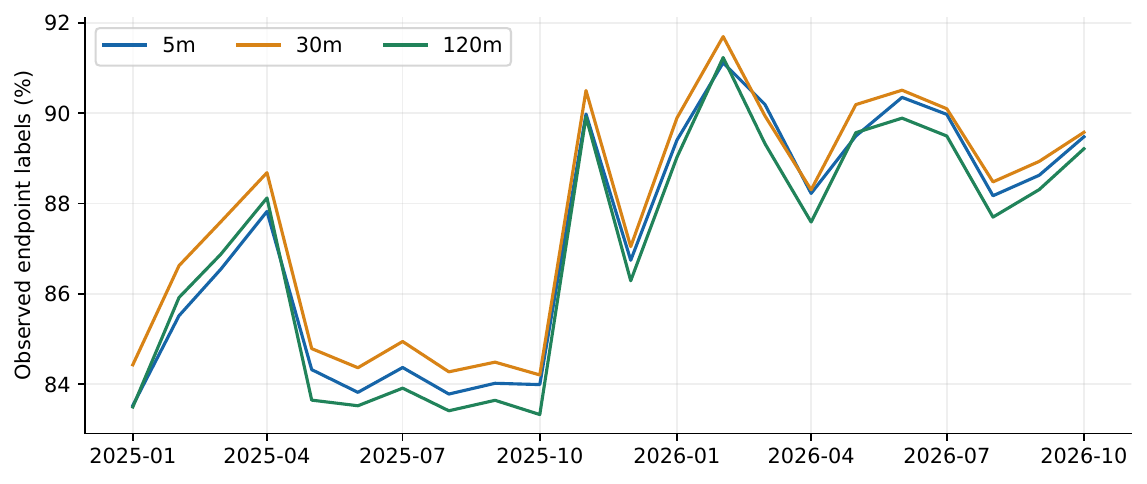}
\caption{Monthly observed total-return label coverage for selected horizons. The denominator excludes horizons extending beyond the market close.}
\label{fig:coverage}
\end{figure}

The availability of these targets determines how much of the forecast universe can enter the empirical evaluation. We distinguish calendar eligibility, which is known at the forecast origin, from the subsequent observability of the realized return. Horizons extending beyond the scheduled market close are ineligible and excluded from the coverage denominator. Among eligible observations, missing required endpoint prices can still prevent construction of a return label. Table~\ref{tab:label-coverage} summarizes coverage by horizon across all 22 test months, counting each observation once rather than repeating it across model variants. Figure~\ref{fig:coverage} complements these aggregate counts by showing how coverage varies across months for selected horizons. Across the full evaluation period, the 30-minute label has approximately 87.64\% coverage.

\begin{table}[H]\centering
\begin{tabular}{lrrr}
\toprule
Horizon & Time-eligible rows & Observed total & Coverage (\%) \\
\midrule
5m & 5,248,638 & 4,576,264 & 87.19 \\
10m & 5,248,638 & 4,573,370 & 87.13 \\
15m & 5,248,638 & 4,575,670 & 87.18 \\
20m & 5,248,638 & 4,577,204 & 87.21 \\
25m & 5,248,638 & 4,579,273 & 87.25 \\
30m & 5,248,638 & 4,600,076 & 87.64 \\
60m & 4,809,788 & 4,197,752 & 87.28 \\
120m & 3,932,088 & 3,417,086 & 86.90 \\
\bottomrule
\end{tabular}

\caption{Endpoint-label coverage across all 22 test months; counted once, not twelve times across variants.}
\label{tab:label-coverage}
\end{table}


\section{Transformer architecture and training}
\label{sec:architecture}
Transformers construct contextual representations through self-attention where each input learns how strongly to combine information from the other inputs. The original architecture combines multi-head attention with feedforward transformations, residual connections and layer normalization \cite{vaswani2017attention}. GPT-style language models apply causal self-attention to ordered token sequences for autoregressive prediction \cite{radford2019gpt2}. Here, we use a cross-sectional Transformer for return regression where the inputs are stocks observed at the same forecast origin and attention captures relationships among their feature representations. Historical information enters through engineered features rather than through temporal attention over a sequence of raw minute bars.

Let $t$ denote a forecast-origin timestamp on the 30-minute grid, and let $N_t$ be the number of stocks with model inputs at that timestamp. The index $i=1,\ldots,N_t$ identifies a stock within that cross-section. A single snapshot is represented by $X_t\in\mathbb R^{N_t\times572}$. A mini-batch contains $B=16$ such snapshots. Because $N_t$ can vary across origins, padding gives a tensor of shape $[B,N_{\max},572]$, where $N_{\max}$ is the largest stock count in that batch. Attention operates within each snapshot along the stock dimension, independently of the other batched origins. Padded stocks are excluded by an attention mask and no causal ordering is imposed across stocks, since their features belong to the same information set at time $t$. A shared linear projection maps each stock's 572 features to a 32-dimensional representation. One pre-norm Transformer layer applies four-head self-attention and a stock-wise feedforward transformation with hidden width 64 and GELU activation. Layer normalization is applied before each attention and feedforward sublayer, followed by the corresponding residual addition. For each attention head, learned query, key and value projections produce the operation
\[
\operatorname{Attention}(Q,K,V)=\operatorname{softmax}\!\left(\frac{QK^\top}{\sqrt{d_k}}+A\right)V,
\]
where $d_k=8$ is the per-head query/key dimension and $A$ masks padded key positions. The softmax is taken over stocks within the snapshot while the four head outputs are concatenated and projected back to width 32. Following the attention/feedforward processing and normalization, a shared prediction head maps each stock's representation to eight horizon-specific outputs. The unpadded output for origin $t$ therefore has shape $[N_t,8]$, with entry $(i,h)$ predicting that stock's scaled residual return at horizon $h$. Table~\ref{tab:model-settings} summarizes the architecture and optimization settings, which are held fixed across historical-sampling policies.

\begin{table}[htbp]\centering
\renewcommand{\arraystretch}{1.1}
\begin{tabular}{lr}\toprule
Setting & Value\\\midrule
Input features & 572\\
Projected model width & 32\\
Attention heads & 4\\
Transformer layers & 1\\
Feedforward activation & GELU\\
Normalization & Pre-norm\\
Feedforward width & 64\\
Dropout & 0.10\\
Output horizons & 8\\
Batch size & 16 snapshots\\
Optimizer & AdamW\\
Learning rate & $5\times10^{-4}$\\
Weight decay & $10^{-4}$\\
Gradient norm clipping & 1.0\\
Maximum selection epochs / patience & 20 / 2\\
Initialization seed & 7\\\bottomrule
\end{tabular}
\caption{Cross-sectional Transformer architecture and training settings shared by all twelve historical-selection configurations.}
\label{tab:model-settings}
\end{table}

\subsection{Scaling, masking and loss function}
Preprocessing and the training objective use only information from the permitted fitting sample. Feature means and standard deviations are fitted using the selection-training months. Features are then standardized, non-finite standardized entries are replaced with zero (or clipped for infinities), and values are clipped to $[-10,10]$. These are preprocessing safeguards rather than learned winsorization quantiles. Asset padding is masked in attention while missing target labels are masked in the loss. A missing target therefore does not remove an otherwise present asset from the attention input.

For each horizon, $s_h$ is the standard deviation of finite residual-return labels expressed in basis points and pooled across stocks and timestamps in the permitted fitting sample $\mathcal F$:
\begin{equation}
s_h=\operatorname{Std}_{(i,t)\in\mathcal F}\!\left(10^4y_{i,t,h}\right),\qquad
\widetilde y_{i,t,h}=\frac{10^4y_{i,t,h}}{s_h},\qquad
\widehat y_{i,t,h}=\frac{s_h\widehat{\widetilde y}_{i,t,h}}{10^4}.
\end{equation}
The labels are scaled without subtracting their mean. Each fitted model uses one fixed scale per horizon, shared across stocks and forecast origins. For example, $s_h=20$ basis points and a scaled prediction of $0.5$ correspond to a predicted residual return of 10 basis points. Selection fits estimate these scales using only selection-training observations. A fresh refit recomputes feature and target scalers on its train-plus-validation sample before applying them to the subsequent test period.

The objective combines prediction accuracy with cross-sectional association. Define the scaled error as $e_{i,t,h}=\widehat{\widetilde y}_{i,t,h}-\widetilde y_{i,t,h}$. The error penalty is the Huber loss with threshold one,
\begin{equation}
\operatorname{Huber}_{1}(e)=
\begin{cases}
\tfrac12e^2, & |e|\leq1,\\
|e|-\tfrac12, & |e|>1.
\end{cases}
\end{equation}
The subscript $1$ specifies the transition threshold in scaled-target units, not an observation index. Small errors receive a quadratic penalty while larger errors receive a linear penalty that limits their influence relative to squared error. At threshold one, this expression equals the smooth-L1 loss used in the implementation.

For a mini-batch, let $M_{i,t,h}=1$ when a stock has a finite target and $0$ otherwise. Let $\mathcal H_B$ contain the horizons with at least one observed target. The loss is
\begin{equation}
\mathcal L=\frac{1}{|\mathcal H_B|}\sum_{h\in\mathcal H_B}
\left[
\frac{\sum_{t,i}M_{i,t,h}\operatorname{Huber}_{1}(e_{i,t,h})}{\sum_{t,i}M_{i,t,h}}
-0.25\,\overline{\operatorname{Corr}}_h
\right].
\end{equation}
Here $\operatorname{Corr}_{t,h}$ measures Pearson correlation between predictions and targets across observed stocks at a fixed origin $t$ and horizon $h$. To state the implemented calculation precisely, let $p_i=\widehat{\widetilde y}_{i,t,h}$ and $z_i=\widetilde y_{i,t,h}$, with $\bar p$ and $\bar z$ their means over $\mathcal O_{t,h}=\{i:M_{i,t,h}=1\}$. Then
\begin{equation}
\operatorname{Corr}_{t,h}=
\frac{\sum_{i\in\mathcal O_{t,h}}(p_i-\bar p)(z_i-\bar z)}
{\sqrt{\max\!\left\{\left[\sum_{i\in\mathcal O_{t,h}}(p_i-\bar p)^2\right]
\left[\sum_{i\in\mathcal O_{t,h}}(z_i-\bar z)^2\right],10^{-8}\right\}}}.
\end{equation}
The lower bound inside the square root is a numerical safeguard. Away from this bound, the expression is the usual Pearson correlation. Let $\mathcal T_{B,h}$ be the batch origins with at least five observed labels at horizon $h$. The horizon-level correlation reward is
\[
\overline{\operatorname{Corr}}_h=
\frac{\sum_{t\in\mathcal T_{B,h}}\operatorname{Corr}_{t,h}}
{\max\{1,|\mathcal T_{B,h}|\}},
\]
which is zero when no origin meets the five-label minimum. Subtracting this term rewards higher cross-sectional correlation. The objective first aggregates each horizon's error and correlation terms and then weights active horizons equally. Reported Rank IC uses Spearman correlation, so it is distinct from the Pearson term optimized during training.

\subsection{Early stopping, fresh refit and chronology}
One training example is a complete market snapshot containing the stocks observed at a single forecast time. At the start of each epoch, these snapshots are placed in a new random order and grouped into mini-batches. The stocks belonging to a snapshot stay together so the model can attend across them. This shuffling changes the order in which training examples are processed while the chronological assignment of months to training, validation and testing remains unchanged.

Validation uses the same Huber-minus-$0.25$-Pearson-correlation criterion as training. A selection fit stops after two epochs without an improvement in validation loss, and the best epoch can therefore precede the last completed epoch. The test period is not used for this choice.

The final model starts again from the saved initialization, uses a separate deterministic dropout stream, and trains on the permitted refit months for the selected number of epochs. The next month supplies the out-of-sample test. Labels stay within the same trading session and the selection and test boundaries are monthly. A label therefore cannot cross the month boundary into the next split.

Table~\ref{tab:training-duration} summarizes the selected training durations across policies.
\begin{table}[H]\centering\small
\begin{tabular}{lrrrr}
\toprule
Model & Mean refit epochs & Median & Max & Selection epochs \\
\midrule
C6 & 4.32 & 4.0 & 7 & 6.32 \\
C12 & 5.23 & 5.5 & 11 & 7.23 \\
C24 & 2.95 & 3.0 & 6 & 4.95 \\
U6 & 4.14 & 4.0 & 9 & 6.14 \\
U12 & 3.73 & 3.5 & 7 & 5.73 \\
U24 & 3.36 & 3.0 & 7 & 5.36 \\
R6 & 4.50 & 4.0 & 11 & 6.50 \\
R12 & 3.64 & 3.0 & 12 & 5.64 \\
R24 & 3.64 & 3.0 & 7 & 5.64 \\
M6 & 4.23 & 4.0 & 6 & 19.32 \\
M12 & 3.91 & 4.0 & 9 & 17.68 \\
M24 & 3.55 & 3.5 & 6 & 17.00 \\
\bottomrule
\end{tabular}

\caption{Training duration selected per final model. Selection epochs sum all validation fits. M uses three fits, so its total is not directly comparable to a single fit.}
\label{tab:training-duration}
\end{table}
The selected refit count has median 4, minimum 1 and maximum 12. Only 59 of 264 final models select more than five epochs. No selection fit reaches the twenty-epoch cap.

We compare predictive performance across twelve historical-selection configurations and eight forecast horizons using the metrics defined in Section~\ref{sec:evaluation}. We then examine how rankings change with training quantity and calendar period, assess selected pairwise differences and interpret forecast calibration and diversity. All comparisons concern observed-label forecast accuracy rather than realized trading profitability.\section{Results}
\label{sec:results}

We first compare predictive performance across the twelve historical-selection policies and eight forecast horizons, using the metrics which we define next in  sub section~\ref{sec:evaluation}. We then examine how the rankings change with training quantity and calendar period, assess selected pairwise differences and interpret forecast calibration and diversity. The final subsection brings the supported findings and their limitations together. All comparisons concern observed-label forecast accuracy rather than realized trading profitability.

\subsection{Evaluation methodology}
\label{sec:evaluation}
We evaluate the monthly out-of-sample forecasts along three complementary dimensions: their association with realized residual returns, the magnitude and stability of their predictions, and the uncertainty in comparisons between historical-selection policies. Information coefficients provide the main accuracy measure. An equal-weight forecast ensemble supplies a common combination benchmark, while calibration, dependence and persistence diagnostics help interpret what the accuracy results imply. The following definitions underpin the comparisons reported in Section~\ref{sec:results}.

\subsubsection{Information coefficients}
At each origin $t$ and horizon $h$, on the common observed target set $\mathcal O_{t,h}$,
\begin{align}
IC^{P}_{t,h}&=\operatorname{Corr}_{i\in\mathcal O_{t,h}}(\widehat y_{i,t,h},y_{i,t,h}),\\
IC^{S}_{t,h}&=\operatorname{Corr}_{i\in\mathcal O_{t,h}}
\bigl(\operatorname{rank}(\widehat y_{i,t,h}),\operatorname{rank}(y_{i,t,h})\bigr).
\end{align}
Here $\operatorname{Corr}$ denotes Pearson correlation, so $IC^S$ is Spearman correlation obtained by applying Pearson correlation to ranks. Ranks use average ranks for ties while evaluation requires at least twenty valid stocks and nonzero cross-sectional variance. Daily IC is the mean across eligible origins and period IC is the mean across daily ICs. Thus a session with fewer origins is not automatically down-weighted. Tables that follow display correlations multiplied by 100 so an IC of $1.90\%$ means a correlation of $0.0190$, not a forecast return of $1.90\%$. All configurations have matched stock--timestamp prediction keys and share the underlying labels and calendar mask. We verified that training and refitting preceded each test period and inspected the timestamps used to construct return labels. Independently recalculated Pearson and Spearman correlations matched the saved metrics for selected observations at the $5$-, $30$- and $120$-minute horizons.

\subsubsection{Equal forecast ensemble}
For each stock $i$, forecast origin $t$ and horizon $h$, the ensemble takes the arithmetic mean of all twelve models' predictions, with equal weight $1/12$ on each model. The members are C6, C12, C24, U6, U12, U24, R6, R12, R24, M6, M12 and M24, using the model refitted for the corresponding test month. The index $m$ below identifies these twelve policy--quantity configurations. Predictions are combined at the same stock, time and horizon, without averaging across stocks, origins or horizons. The ensemble forecast is formed before its IC is computed:
\begin{equation}
\widehat y^{E}_{i,t,h}=\frac{1}{12}\sum_{m=1}^{12}\widehat y^{(m)}_{i,t,h}.
\end{equation}
It averages decimal-return forecasts, not ranks or IC values, and uses no outcome-fitted model weights. This diagnostic ensemble was added after the training experiment. Its evaluation remains out of sample at the model level, while comparisons and policy selection using the full test period are retrospective.

\subsubsection{Calibration, dependence and persistence}
\label{sec:diagnostic-methods}
These diagnostics address questions that IC alone cannot answer: whether forecast magnitudes align with realized returns, whether different models contribute distinct forecasts, and how stable predictions remain between adjacent origins. Their results are reported in Section~\ref{sec:calibration-results}. For each horizon, pooled observed-label out-of-sample calibration estimates
\begin{equation}y_{i,t,h}=\alpha_h+\beta_h\widehat y^E_{i,t,h}+\epsilon_{i,t,h}.\end{equation}
An intercept of zero and a slope of one would indicate unit linear calibration. These estimated slopes and intercepts are descriptive and were not used to rescale forecasts in this study. Pooling all stock--origins weights larger observed cross-sections more heavily than equal-day IC and can be affected by extreme labels.

Between-model dependence is the average within-origin Pearson correlation of predictions and measures similarity between models. Forecast persistence is the cross-sectional correlation of a model's predictions across adjacent 30-minute origins, matching stock names and excluding overnight pairs. It measures stability between successive forecast cross-sections rather than a per-stock long-run time-series autocorrelation. Sign-flip rates use the same matched pairs. Their reported aggregates average monthly diagnostic means.

\subsubsection{Exploratory uncertainty}
For selected sampling-versus-contiguous comparisons, we compute a paired difference in monthly mean IC and resample the $22$ paired months $10,000$ times. Percentile intervals summarize uncertainty in the equally weighted monthly mean difference. These differ slightly from the day-weighted full-period IC differences in the main tables. The intervals are exploratory because months are resampled as independent units despite shared rolling histories and possible serial dependence. No multiple-comparison correction or alternative seeds are included. A positive point estimate or interval is not sufficient to choose a production policy. The strongest model at each horizon was selected after observing the same test sample.

\subsection{Full-period predictive performance}
\begin{figure}[htbp]\centering
\includegraphics[width=0.75\textwidth]{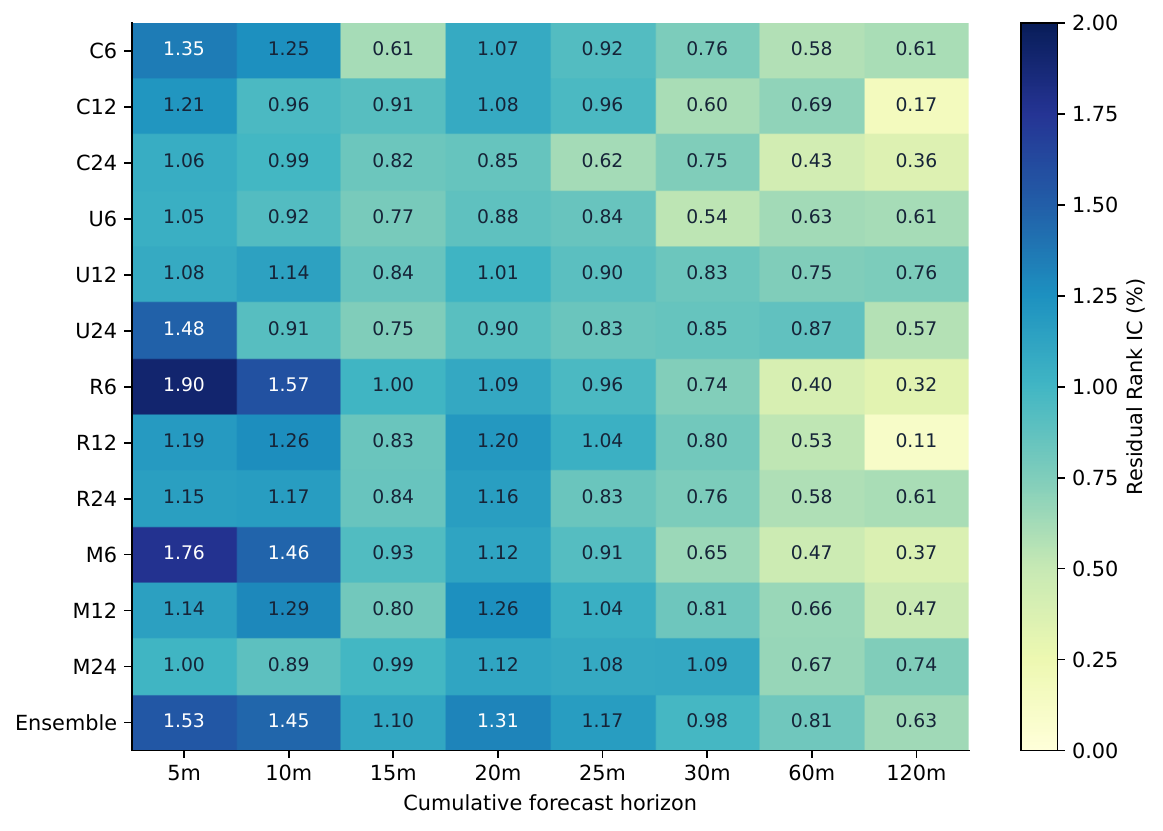}
\caption{Observed-label Rank IC across all configurations and horizons.}
\label{fig:ic_heatmap}
\end{figure}

Table~\ref{tab:rank-performance} and Figure~\ref{fig:ic_heatmap} summarize the full-period results while Table~\ref{tab:best-horizon} compares the retrospectively best individual model with the equal ensemble at each horizon. Full-period Pearson ICs are provided for comparison in \ref{app:additional-ic}, Table~\ref{tab:appendix-pearson}. The central finding is that the leading historical-selection configuration depends on the forecast horizon: R6 leads individual models at 5--15 minutes, M12 leads at 20 minutes, M24 at 25--30 minutes, U24 at 60 minutes and U12 at 120 minutes. Thus no single policy or training quantity dominates across horizons.

At 5 and 10 minutes, R6 achieves Rank ICs of 1.90\% and 1.57\%, compared with 1.35\% and 1.25\% for C6. These differences show that equal nominal training quantities can yield materially different point estimates when the selected history changes. R6 samples with a recency preference from a broader historical pool whereas C6 uses the latest contiguous months. The result therefore suggests value in selective use of historical observations rather than establishing that the most recent observations alone are sufficient. Six-month training leads within C, R and M at 5 minutes, but U24 leads within U. Even at the shortest horizon, a preference for less training data is conditional on the sampling policy.

The leading individual configurations use 12 or 24 training months from 20 minutes onward. However, U12 exceeds U24 at 120 minutes and C24 is never the leading contiguous configuration. Longer horizons therefore do not imply a uniformly increasing requirement for training data. M's performance at 20--30 minutes must also be interpreted in light of its different validation procedure.

\begin{table}[H]\centering\small\begin{tabular}{lrrrrrrrr}
\toprule
Model & 5m & 10m & 15m & 20m & 25m & 30m & 60m & 120m \\
\midrule
C6 & 1.35 & 1.25 & 0.61 & 1.07 & 0.92 & 0.76 & 0.58 & 0.61 \\
C12 & 1.21 & 0.96 & 0.91 & 1.08 & 0.96 & 0.60 & 0.69 & 0.17 \\
C24 & 1.06 & 0.99 & 0.82 & 0.85 & 0.62 & 0.75 & 0.43 & 0.36 \\
U6 & 1.05 & 0.92 & 0.77 & 0.88 & 0.84 & 0.54 & 0.63 & 0.61 \\
U12 & 1.08 & 1.14 & 0.84 & 1.01 & 0.90 & 0.83 & 0.75 & \textbf{0.76} \\
U24 & 1.48 & 0.91 & 0.75 & 0.90 & 0.83 & 0.85 & \textbf{0.87} & 0.57 \\
R6 & \textbf{1.90} & \textbf{1.57} & 1.00 & 1.09 & 0.96 & 0.74 & 0.40 & 0.32 \\
R12 & 1.19 & 1.26 & 0.83 & 1.20 & 1.04 & 0.80 & 0.53 & 0.11 \\
R24 & 1.15 & 1.17 & 0.84 & 1.16 & 0.83 & 0.76 & 0.58 & 0.61 \\
M6 & 1.76 & 1.46 & 0.93 & 1.12 & 0.91 & 0.65 & 0.47 & 0.37 \\
M12 & 1.14 & 1.29 & 0.80 & 1.26 & 1.04 & 0.81 & 0.66 & 0.47 \\
M24 & 1.00 & 0.89 & 0.99 & 1.12 & 1.08 & \textbf{1.09} & 0.67 & 0.74 \\
Ensemble & 1.53 & 1.45 & \textbf{1.10} & \textbf{1.31} & \textbf{1.17} & 0.98 & 0.81 & 0.63 \\
\bottomrule
\end{tabular}

\caption{Mean observed-label residual Rank IC (\%). Bold denotes the largest point estimate in each column, including the ensemble. C/U/R/M denote contiguous/uniform/recency/multi-recency, and the number denotes selected training months.}\label{tab:rank-performance}\end{table}
\begin{table}[H]\centering\begin{tabular}{lrrr}
\toprule
Horizon & Best individual & Rank IC (\%) & Ensemble IC (\%) \\
\midrule
5m & R6 & 1.90 & 1.53 \\
10m & R6 & 1.57 & 1.45 \\
15m & R6 & 1.00 & 1.10 \\
20m & M12 & 1.26 & 1.31 \\
25m & M24 & 1.08 & 1.17 \\
30m & M24 & 1.09 & 0.98 \\
60m & U24 & 0.87 & 0.81 \\
120m & U12 & 0.76 & 0.63 \\
\bottomrule
\end{tabular}

\caption{Retrospective best individual model versus the equal forecast ensemble.}\label{tab:best-horizon}\end{table}

Averaging all twelve models exceeds the best individual mean at 15, 20 and 25 minutes, with ensemble ICs of 1.10\%, 1.31\% and 1.17\%, respectively. The corresponding gains are 0.10, 0.05 and 0.09 percentage points of IC. At the other five horizons, including 30 minutes, at least one individual model performs better. Equal averaging therefore provides a useful combination benchmark but can dilute the advantage of a stronger individual model. These comparisons do not establish statistically significant ensemble gains and we note that the best individual at each horizon is identified retrospectively rather than selected using information available before the test period.

Although IC is generally higher at the shortest horizons than at 60--120 minutes, it does not decline monotonically. For example, the ensemble rises from 1.10\% at 15 minutes to 1.31\% at 20 minutes. Because the targets are cumulative returns over different horizons, these comparisons do not directly estimate the decay of a fixed trading signal. Taken together, the results support treating historical selection and training quantity as horizon-dependent modelling choices. The uncertainty and calendar-period analyses below assess how strongly these full-period rankings can be generalized.

\subsection{Training quantity within each policy family}
\begin{figure}[htbp]\centering
\includegraphics[width=0.80\textwidth]{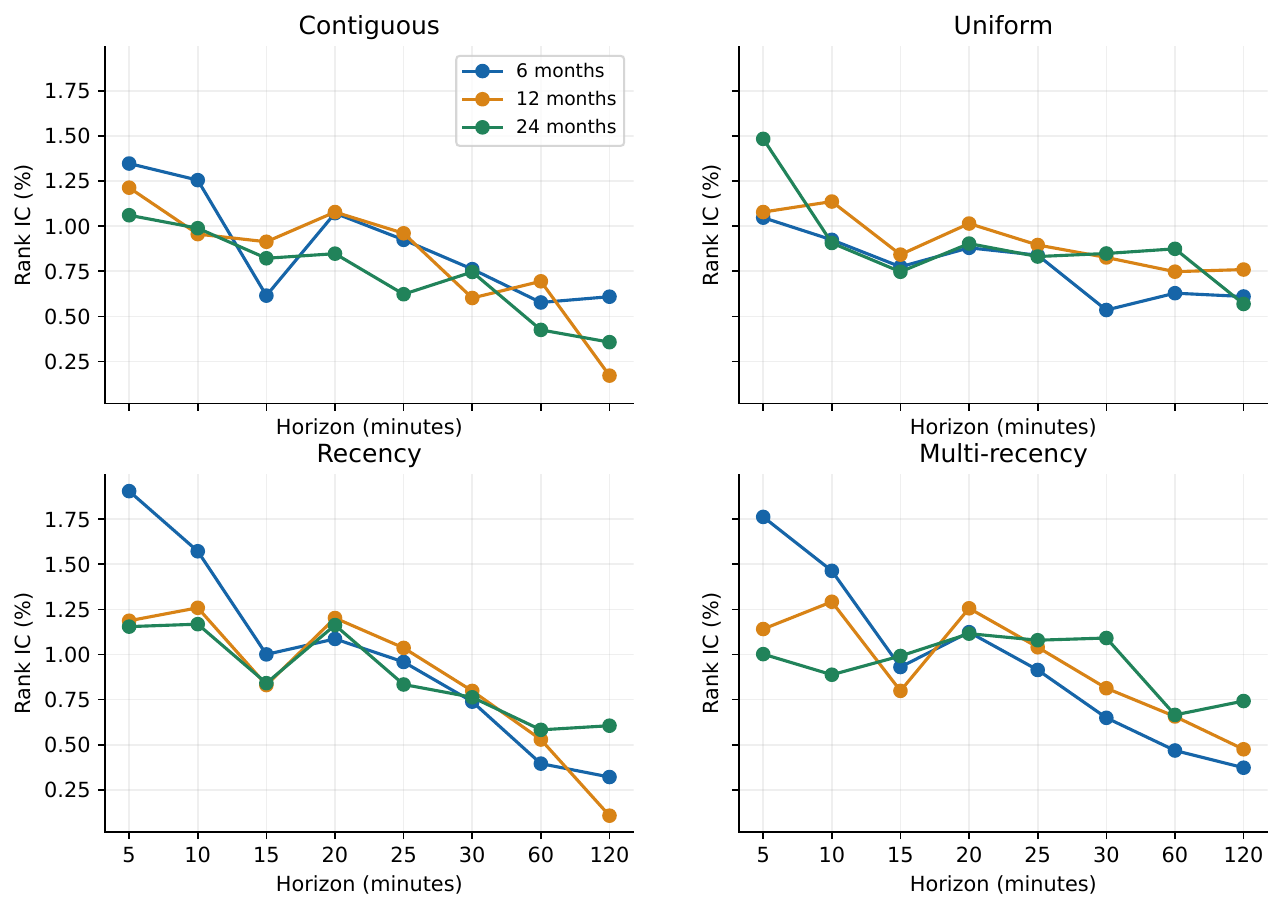}
\caption{Training quantity within each policy. Each panel retains its policy's validation procedure while historical pool breadth changes with quantity for sampled methods.}
\label{fig:quantity_horizon}
\end{figure}

Table~\ref{tab:quantity-winners} identifies the leading training quantity within each policy and Figure~\ref{fig:quantity_horizon} shows how performance changes across horizons.
\begin{table}[H]\centering\small\begin{tabular}{lrrrrrrrr}
\toprule
Sampling & 5m & 10m & 15m & 20m & 25m & 30m & 60m & 120m \\
\midrule
Contiguous & 6 & 6 & 12 & 12 & 12 & 6 & 12 & 6 \\
Uniform & 24 & 12 & 12 & 12 & 12 & 24 & 24 & 12 \\
Recency & 6 & 6 & 6 & 12 & 12 & 12 & 24 & 24 \\
Multi-recency & 6 & 6 & 24 & 12 & 24 & 24 & 24 & 24 \\
\bottomrule
\end{tabular}

\caption{Training-month quantity with the largest mean Rank IC within each sampling policy. Some differences are small, so these maxima should not be interpreted as significance tests.}\label{tab:quantity-winners}\end{table}
The clearest monotone horizon pattern appears within R: 6 months win at 5--15 minutes, 12 at 20--30 minutes and 24 at 60--120 minutes. Under M, 24 months generally help the longer horizons while six months dominate at 5--10 minutes. C and U do not follow a simple monotone rule. In particular, contiguous 24-month training wins none of the eight columns within its family while uniform 24-month training wins the 5-, 30- and 60-minute columns.

These observations support testing the hypothesis that short-horizon predictability is more dependent on recent conditions. These within-family comparisons retain the same sampling rule and validation procedure, although sampled policies also change historical pool breadth with $T$. At the same nominal $T$, contiguous training uses only the latest months while recency sampling can include older months. Their comparison therefore concerns the composition of the training history rather than simply newer versus older data.

\subsection{Stability across periods}
\begin{figure}[htbp]\centering
\includegraphics[width=0.78\textwidth]{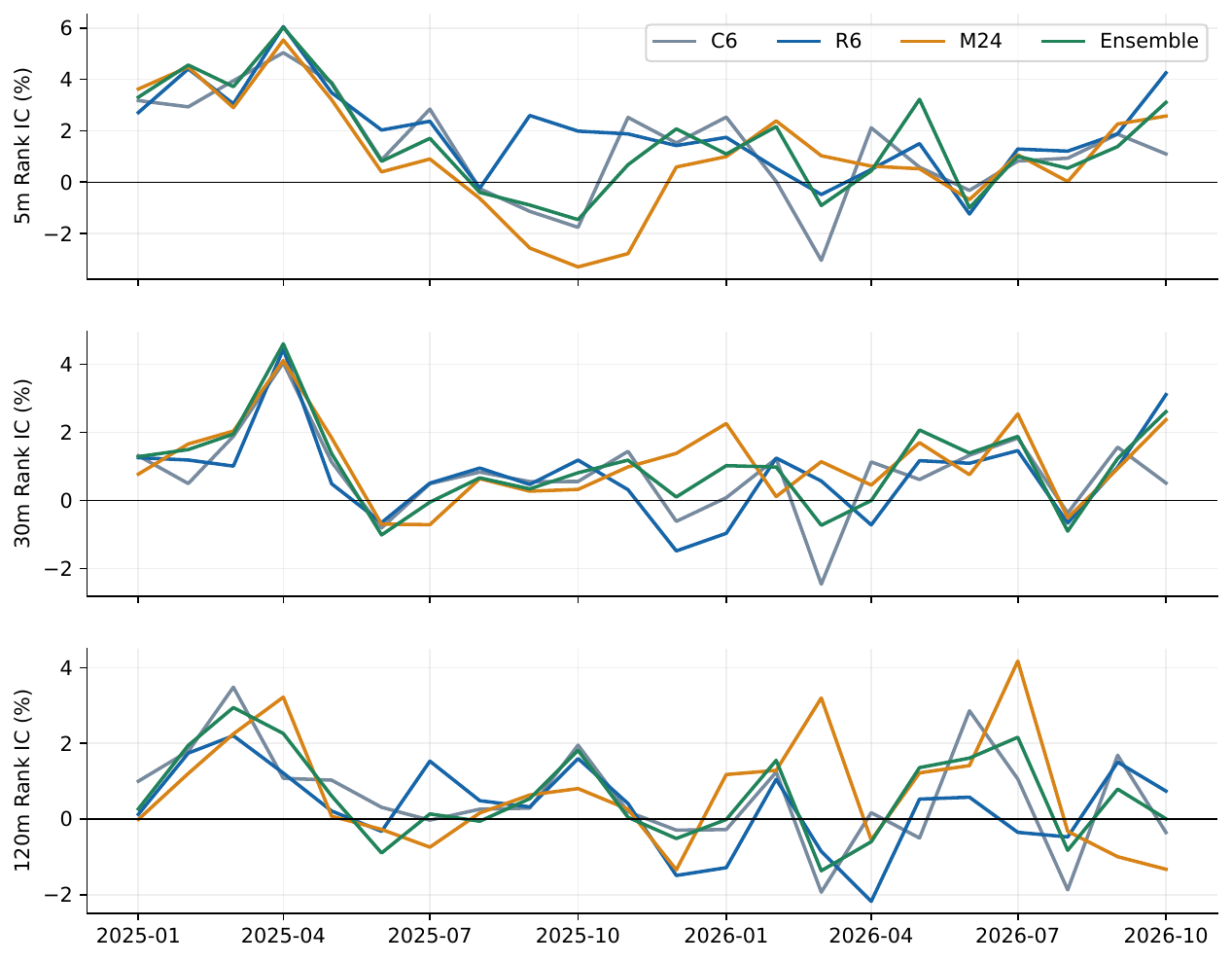}
\caption{Monthly Rank IC for C6, R6, M24 and the equal ensemble. Monthly variation is substantial.}
\label{fig:monthly_ic}
\end{figure}

Table~\ref{tab:ensemble-periods} compares the ensemble across subperiods and Figure~\ref{fig:monthly_ic} displays the monthly paths of selected policies and the ensemble. For reference, \ref{app:additional-ic} reports all configurations separately for 2025 and 2026 in Tables~\ref{tab:appendix-2025} and~\ref{tab:appendix-2026}, with the two halves of 2025 detailed in Tables~\ref{tab:appendix-2025h1} and~\ref{tab:appendix-2025h2}.
\begin{table}[H]\centering\begin{tabular}{lrrrr}
\toprule
Horizon & H1 2025 & H2 2025 & 2026 & All \\
\midrule
5m & 3.72 & 0.28 & 1.01 & 1.53 \\
10m & 2.43 & 1.44 & 0.88 & 1.45 \\
15m & 1.49 & 1.23 & 0.78 & 1.10 \\
20m & 1.72 & 1.18 & 1.14 & 1.31 \\
25m & 1.66 & 0.84 & 1.09 & 1.17 \\
30m & 1.65 & 0.50 & 0.89 & 0.98 \\
60m & 1.32 & 0.54 & 0.67 & 0.81 \\
120m & 1.20 & 0.35 & 0.47 & 0.63 \\
\bottomrule
\end{tabular}

\caption{Equal-ensemble Rank IC (\%) across subperiods. 2026 ends on 20 October.}\label{tab:ensemble-periods}\end{table}
The deterioration is especially sharp at 5 minutes and remains visible at 30 minutes. Changes in market conditions are a possible explanation, but this experiment does not identify the cause of the decline. The observed variation motivates evaluating whether alternative update schedules or feature revisions improve performance in subsequent periods. It does not establish that either intervention would resolve the deterioration.

\subsection{How strong is the relative evidence?}
Table~\ref{tab:uncertainty} examines whether selected average performance differences are consistently supported across the 22 test months. For each comparison a positive gain means that the sampled-history model has a higher monthly mean Rank IC than its contiguous baseline. The bootstrap interval describes uncertainty in the average of these paired monthly differences. An interval containing zero means that the calculation does not resolve the direction of the average difference at the stated confidence level. It does not establish that the models are equivalent.
\begin{table}[H]\centering\small\begin{tabular}{lrrrr}
\toprule
Comparison & Horizon & Mean gain (pp) & 95\% interval (pp) & Months won \\
\midrule
R6 vs C6 & 5m & 0.579 & [-0.011, 1.226] & 13/22 \\
R6 vs C6 & 10m & 0.334 & [-0.067, 0.731] & 14/22 \\
R6 vs C6 & 30m & 0.008 & [-0.413, 0.489] & 10/22 \\
M24 vs C24 & 30m & 0.325 & [0.028, 0.632] & 14/22 \\
M24 vs C24 & 120m & 0.364 & [-0.201, 1.083] & 9/22 \\
\bottomrule
\end{tabular}

\caption{Exploratory paired-month bootstrap comparisons. Gains and intervals are percentage points of IC, not returns. No multiple-testing adjustment is applied.}\label{tab:uncertainty}\end{table}
R6 exceeds C6 at 5 minutes in 13 of 22 months, with an estimated mean gain of 0.579 percentage points and a 95\% interval of $[-0.011,1.226]$. These results favour R6 on average in this sample. However, the interval does not establish a reliably positive advantage. At 30 minutes, the estimated R6--C6 gain is just 0.008 percentage points, so the two methods have almost identical average monthly Rank IC in the observed sample. This similarity concerns the aggregate metric rather than the individual forecasts or their month-by-month performance. The 95\% interval of $[-0.413,0.489]$ percentage points still allows advantages in either direction and does not establish statistical equivalence.

M24 exceeds C24 at 30 minutes by an estimated 0.325 percentage points, with an interval of $[0.028,0.632]$ and wins in 14 of 22 months. This is the only reported comparison whose interval lies entirely above zero and it provides stronger evidence within this exploratory calculation. M24's 120-minute interval includes zero despite a positive estimated mean gain and we also highlight that M uses a different validation procedure from C, so their difference cannot be attributed solely to historical sampling.

\subsection{Calibration and ensemble diversity}
\label{sec:calibration-results}
Rank IC measures whether forecasts correctly order stocks. It does not tell us whether their numerical magnitudes are accurate. It also does not show how similar the twelve models are or how much their forecasts change between successive origins. We address these three questions using the diagnostics defined in Section~\ref{sec:diagnostic-methods}.

\begin{figure}[htbp]\centering
\includegraphics[width=0.70\textwidth]{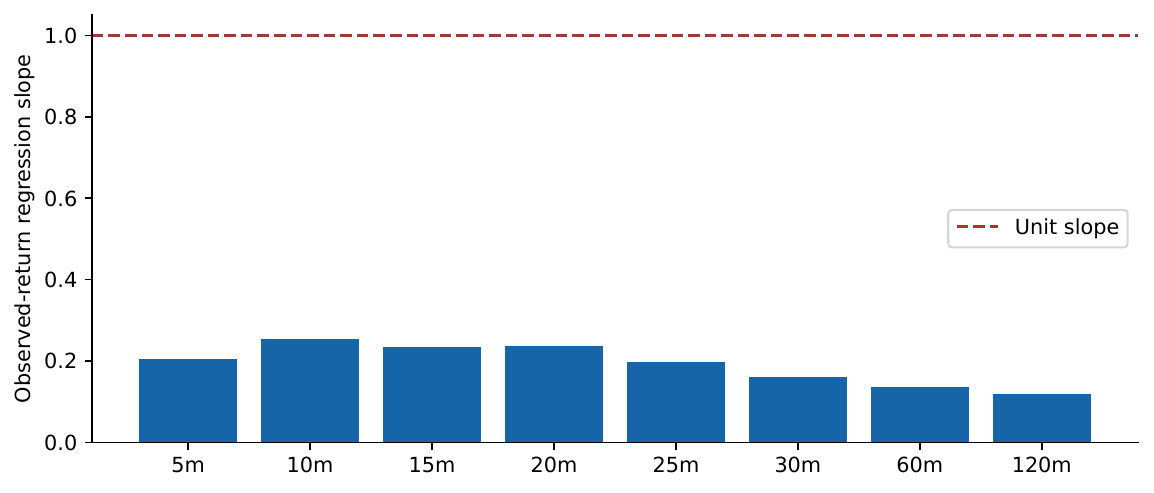}
\caption{Realized residual returns regressed on ensemble forecasts. Unit linear calibration would have slope one and intercept zero, while the estimated slopes are substantially below one.}
\label{fig:calibration}
\end{figure}

Table~\ref{tab:calibration} and Figure~\ref{fig:calibration} examine forecast magnitude by regressing realized residual returns on the ensemble forecasts. In this regression, a slope of one and an intercept of zero would indicate unit linear calibration. The estimated slopes of approximately 0.12--0.25 mean that differences in forecasts correspond to substantially smaller differences in fitted realized returns. For example, at a slope of 0.20, a 10-basis-point difference between two forecasts corresponds to a 2-basis-point difference in the regression's fitted outcomes. This is a statement about the pooled fitted relationship, not a guarantee for an individual stock or prediction.
\begin{table}[H]\centering\small\begin{tabular}{lrrrr}
\toprule
Horizon & Slope & Intercept (bp) & Forecast SD (bp) & Label SD (bp) \\
\midrule
5m & 0.206 & -0.006 & 1.21 & 21.97 \\
10m & 0.255 & 0.004 & 1.47 & 29.68 \\
15m & 0.235 & 0.016 & 1.53 & 35.43 \\
20m & 0.236 & 0.007 & 1.83 & 40.12 \\
25m & 0.197 & 0.022 & 2.02 & 44.56 \\
30m & 0.162 & 0.042 & 2.07 & 48.24 \\
60m & 0.136 & 0.032 & 3.05 & 65.43 \\
120m & 0.118 & 0.072 & 4.34 & 88.81 \\
\bottomrule
\end{tabular}

\caption{Pooled observed-label calibration of the equal ensemble. Slopes were estimated for diagnosis only.}\label{tab:calibration}\end{table}
The forecasts can therefore contain useful ranking information while their raw magnitudes should not be interpreted as calibrated expected returns. The much larger volatility of realized returns is a separate issue, because realized outcomes include unpredictable noise and need not have the same variance as forecasts. These regressions are descriptive and pool observations across stocks and dates. Any recalibration intended for forecasting must be fitted only on earlier outcomes and tested on subsequent data, rather than estimated from the full test period.

\begin{figure}[htbp]\centering
\includegraphics[width=0.70\textwidth]{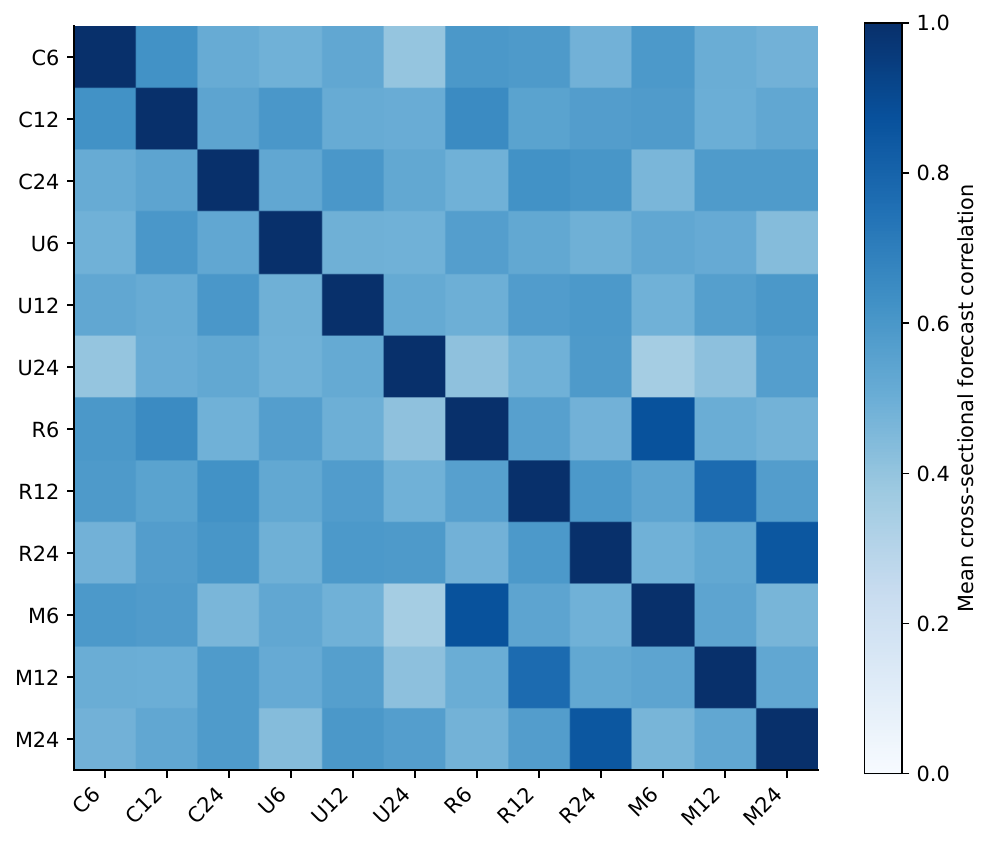}
\caption{Pairwise 30-minute forecast correlations, averaged within origins and then across months. Shared architecture and inputs do not produce identical forecasts.}
\label{fig:correlations_30m}
\end{figure}

Table~\ref{tab:persistence} and Figure~\ref{fig:correlations_30m} distinguish similarity across models from stability across time. Between-model correlation compares different models' forecasts for the same stocks at the same origin and horizon. Average correlations of about 0.47--0.61 show that the twelve models produce related but non-identical forecasts. This leaves scope for forecast combination but does not establish independent errors or guarantee that averaging improves accuracy.
\begin{table}[H]\centering\small\begin{tabular}{lrrr}
\toprule
Horizon & Between-model corr. & 30m persistence & Sign flips (\%) \\
\midrule
5m & 0.598 & 0.434 & 34.0 \\
10m & 0.613 & 0.479 & 32.2 \\
15m & 0.573 & 0.474 & 32.3 \\
20m & 0.604 & 0.527 & 31.6 \\
25m & 0.560 & 0.492 & 33.3 \\
30m & 0.542 & 0.457 & 34.6 \\
60m & 0.472 & 0.570 & 31.0 \\
120m & 0.488 & 0.680 & 26.7 \\
\bottomrule
\end{tabular}

\caption{Forecast diversity and stability. Between-model correlation compares different models at the same origin. Persistence compares ensemble forecasts across adjacent 30-minute origins while sign flips report the proportion of matched forecasts that change sign.}\label{tab:persistence}\end{table}
Forecast persistence instead compares the ensemble's cross-sectional forecasts at one origin with its forecasts for the same stocks 30 minutes later, excluding overnight pairs. Persistence is 0.434 at the 5-minute horizon and 0.680 at the 120-minute horizon while the corresponding sign-flip rates are 34.0\% and 26.7\%. Thus the longer-horizon forecasts are more stable between adjacent origins in this sample. Two 120-minute targets issued 30 minutes apart share 90 minutes of their return intervals, whereas two 5-minute targets issued 30 minutes apart do not overlap. That overlap, together with persistent inputs and smoother model responses, can contribute to the observed difference. 

\section{Conclusions}
\label{sec:conclusions}
Expanding an equity dataset to train a deep learning model increases the number of training observations but also introduces older relationships that potentially transfer poorly to current market conditions. The motivation of this study is to examine which parts of the available history are useful for prediction and whether that choice depends on the forecast horizon. We compared twelve historical-selection configurations, combining four policy families with 6-, 12- and 24-month training quantities while holding the cross-sectional Transformer architecture and feature set fixed. A monthly walk-forward experiment produced 264 refits and evaluated eight cumulative return horizons from 5 to 120 minutes. Across these configurations, the leading individual model changes from R6 at 5--15 minutes to M12 at 20 minutes, M24 at 25--30 minutes, U24 at 60 minutes and U12 at 120 minutes. Increasing training quantity does not uniformly improve accuracy and the progression toward larger samples at longer horizons is conditional on the sampling family. These findings support evaluating training quantity, historical selection and forecast horizon jointly, rather than adopting one history length for every target. They also suggest that older observations should be assessed for their predictive contribution rather than retained or discarded solely because of their age.

The equal-weight ensemble of all twelve models exceeds every individual mean at 15, 20 and 25 minutes but does not dominate at the remaining horizons. Historical-selection diversity can therefore be useful for forecast combination, although equal averaging can also dilute a stronger individual forecast. Substantial calendar-period variation further shows why pooled rankings alone are insufficient for selecting a forecasting policy. The calibration diagnostics additionally distinguish useful ranking information from accurately scaled return forecasts. The common-model comparison supports evaluating historical selection jointly with forecast horizon. Its scope is limited by one initialization seed, incomplete exact-endpoint labels and differences in historical pool breadth and validation procedures. Moreover, the exploratory uncertainty intervals do not consistently resolve favourable average differences. The findings therefore motivate prospective comparisons rather than establish a universally optimal policy or a causal benefit from recency. 

Further work could isolate the contribution of recency by fixing training quantity $T$, historical pool $H$, architecture and validation procedure while comparing uniform sampling with recency sampling at different half-lives, such as 1, 3 and 6 months. Equal observation weights after selection would keep this experiment distinct from changing the training loss weights. A complementary comparison could vary $T$ within a fixed pool and sampling rule, recording the resulting age distribution of selected observations. Repeating these experiments across initialization seeds, sampling draws and untouched future periods would help determine whether the horizon-dependent patterns persist. Forecast magnitudes could also be recalibrated using only previously observed outcomes, with subsequent evaluation of sensitivity to outliers and sample coverage. Together, these extensions would clarify whether the observed relationships between training history and forecast horizon provide a reliable basis for improving forecasts.

\newpage
\bibliographystyle{elsarticle-num}
\bibliography{refs}

@article{pesaran2007,
 author = {Pesaran, M. Hashem and Timmermann, Allan},
 title = {Selection of estimation window in the presence of breaks},
 journal = {Journal of Econometrics},
 year = {2007}, volume = {137}, number = {1}, pages = {134--161},
 doi = {10.1016/j.jeconom.2006.03.010}
}

@article{pesaran2013,
 author = {Pesaran, M. Hashem and Pick, Andreas and Pranovich, Mikhail},
 title = {Optimal Forecasts in the Presence of Structural Breaks},
 journal = {Journal of Econometrics},
 year = {2013}, volume = {177}, number = {2}, pages = {134--152},
 doi = {10.1016/j.jeconom.2013.04.002}
}

@article{inoue2017,
 author = {Inoue, Atsushi and Jin, Lu and Rossi, Barbara},
 title = {Rolling window selection for out-of-sample forecasting with time-varying parameters},
 journal = {Journal of Econometrics},
 year = {2017}, volume = {196}, number = {1}, pages = {55--67},
 doi = {10.1016/j.jeconom.2016.03.006}
}

@article{pesaran2011,
 author = {Pesaran, M. Hashem and Pick, Andreas},
 title = {Forecast Combination Across Estimation Windows},
 journal = {Journal of Business \& Economic Statistics},
 year = {2011}, volume = {29}, number = {2}, pages = {307--318},
 doi = {10.1198/jbes.2010.09018}
}

@article{li2022,
 author = {Li, Wendi and Yang, Xiao and Liu, Weiqing and Xia, Yingce and Bian, Jiang},
 title = {{DDG-DA}: Data Distribution Generation for Predictable Concept Drift Adaptation},
 journal = {Proceedings of the AAAI Conference on Artificial Intelligence},
 year = {2022}, volume = {36}, number = {4}, pages = {4092--4100},
 doi = {10.1609/aaai.v36i4.20327}
}

@misc{kaplan2020,
 author = {Kaplan, Jared and McCandlish, Sam and Henighan, Tom and Brown, Tom B. and Chess, Benjamin and Child, Rewon and Gray, Scott and Radford, Alec and Wu, Jeffrey and Amodei, Dario},
 title = {Scaling Laws for Neural Language Models},
 year = {2020}, note = {arXiv:2001.08361},
 doi = {10.48550/arXiv.2001.08361}
}

@inproceedings{zhao2023doubleadapt,
 author = {Zhao, Lifan and Kong, Shuming and Shen, Yanyan},
 title = {{DoubleAdapt}: A Meta-learning Approach to Incremental Learning for Stock Trend Forecasting},
 booktitle = {Proceedings of the 29th ACM SIGKDD Conference on Knowledge Discovery and Data Mining},
 year = {2023}, pages = {3492--3503}, doi = {10.1145/3580305.3599315}
}

@inproceedings{taga2026faf,
 author = {Taga, Ege Onur and Gozeten, Halil Alperen and Tire, Kutay and Dalvi, Rahul and Heckel, Reinhard and Oymak, Samet},
 title = {Filter, Augment, Forecast: Online Data Selection for Robust Time Series Forecasting},
 booktitle = {Proceedings of the 29th International Conference on Artificial Intelligence and Statistics},
 series = {Proceedings of Machine Learning Research},
 year = {2026}, volume = {300}, pages = {1081--1089},
 url = {https://proceedings.mlr.press/v300/taga26a.html}
}

@inproceedings{xie2023dsir,
 author = {Xie, Sang Michael and Santurkar, Shibani and Ma, Tengyu and Liang, Percy},
 title = {Data Selection for Language Models via Importance Resampling},
 booktitle = {Advances in Neural Information Processing Systems},
 year = {2023}, volume = {36},
 pages = {34201--34227}, doi = {10.52202/075280-1482}
}

@inproceedings{xie2023doremi,
 author = {Xie, Sang Michael and Pham, Hieu and Dong, Xuanyi and Du, Nan and Liu, Hanxiao and Lu, Yifeng and Liang, Percy and Le, Quoc V. and Ma, Tengyu and Yu, Adams Wei},
 title = {{DoReMi}: Optimizing Data Mixtures Speeds Up Language Model Pretraining},
 booktitle = {Advances in Neural Information Processing Systems},
 year = {2023}, volume = {36},
 pages = {69798--69818}, doi = {10.52202/075280-3059}
}

@inproceedings{tirumala2023d4,
 author = {Tirumala, Kushal and Simig, Daniel and Aghajanyan, Armen and Morcos, Ari S.},
 title = {{D4}: Improving {LLM} Pretraining via Document De-Duplication and Diversification},
 booktitle = {Advances in Neural Information Processing Systems},
 year = {2023}, volume = {36},
 pages = {53983--53995}, doi = {10.52202/075280-2348}
}

@inproceedings{vaswani2017attention,
  author = {Vaswani, Ashish and Shazeer, Noam and Parmar, Niki and Uszkoreit, Jakob and Jones, Llion and Gomez, Aidan N. and Kaiser, {\L}ukasz and Polosukhin, Illia},
  title = {Attention Is All You Need},
  booktitle = {Advances in Neural Information Processing Systems},
  volume = {30},
  pages = {5998--6008},
  year = {2017},
  url = {https://proceedings.neurips.cc/paper/2017/hash/3f5ee243547dee91fbd053c1c4a845aa-Abstract.html}
}

@techreport{radford2019gpt2,
  author = {Radford, Alec and Wu, Jeffrey and Child, Rewon and Luan, David and Amodei, Dario and Sutskever, Ilya},
  title = {Language Models are Unsupervised Multitask Learners},
  institution = {OpenAI},
  year = {2019},
  url = {https://cdn.openai.com/better-language-models/language_models_are_unsupervised_multitask_learners.pdf}
}

\clearpage
\appendix
\clearpage\section{Additional IC tables}
\label{app:additional-ic}
All entries are correlations multiplied by 100 and use the equal-day aggregation defined in Section~\ref{sec:evaluation}. Bold identifies the largest column value within each table and is descriptive only.
\begin{table}[H]\centering\small\begin{tabular}{lrrrrrrrr}
\toprule
Model & 5m & 10m & 15m & 20m & 25m & 30m & 60m & 120m \\
\midrule
C6 & 1.22 & 1.20 & 0.58 & 0.89 & 0.75 & 0.55 & 0.27 & 0.38 \\
C12 & 1.14 & 1.10 & 1.01 & 1.09 & 0.90 & 0.44 & 0.58 & 0.39 \\
C24 & 0.93 & 1.07 & 0.93 & 0.83 & 0.60 & 0.66 & 0.32 & 0.40 \\
U6 & 0.97 & 0.93 & 0.79 & 0.88 & 0.78 & 0.41 & 0.50 & 0.59 \\
U12 & 0.87 & 1.05 & 0.78 & 0.90 & 0.73 & 0.60 & \textbf{0.61} & 0.70 \\
U24 & 1.20 & 0.84 & 0.71 & 0.62 & 0.61 & 0.50 & 0.53 & 0.42 \\
R6 & \textbf{1.77} & \textbf{1.63} & \textbf{1.14} & 1.11 & 0.97 & 0.69 & 0.34 & 0.35 \\
R12 & 1.06 & 1.28 & 0.84 & 1.17 & 0.99 & 0.67 & 0.44 & 0.34 \\
R24 & 0.94 & 1.03 & 0.77 & 0.98 & 0.64 & 0.56 & 0.47 & 0.70 \\
M6 & 1.65 & 1.49 & 1.04 & 1.13 & 0.92 & 0.62 & 0.41 & 0.39 \\
M12 & 0.95 & 1.24 & 0.80 & \textbf{1.23} & 0.95 & 0.69 & 0.60 & 0.67 \\
M24 & 0.85 & 0.87 & 0.91 & 0.97 & 0.88 & \textbf{0.86} & 0.52 & \textbf{0.72} \\
Ensemble & 1.37 & 1.44 & 1.10 & 1.22 & \textbf{1.03} & 0.76 & 0.60 & 0.64 \\
\bottomrule
\end{tabular}
\caption{Full-period Pearson IC (\%).}\label{tab:appendix-pearson}\end{table}
\begin{table}[H]\centering\small\begin{tabular}{lrrrrrrrr}
\toprule
Model & 5m & 10m & 15m & 20m & 25m & 30m & 60m & 120m \\
\midrule
C6 & 1.93 & 1.52 & 0.75 & 1.26 & 0.98 & 0.94 & 0.72 & 0.92 \\
C12 & 1.86 & 1.58 & 1.31 & 1.30 & 1.16 & 0.93 & 0.75 & 0.28 \\
C24 & 1.27 & 1.44 & 1.26 & 1.06 & 0.86 & 0.80 & 0.67 & 0.51 \\
U6 & 1.65 & 1.15 & 0.89 & 1.08 & 1.00 & 0.68 & 0.49 & 0.62 \\
U12 & 1.33 & 1.45 & 1.08 & 1.29 & 1.21 & \textbf{1.10} & \textbf{1.05} & \textbf{1.16} \\
U24 & 1.61 & 1.01 & 0.73 & 0.80 & 0.71 & 0.78 & 0.74 & 0.62 \\
R6 & \textbf{2.63} & 2.10 & 1.30 & 1.30 & 1.02 & 0.81 & 0.54 & 0.67 \\
R12 & 1.78 & 1.75 & 0.98 & 1.22 & 1.21 & 0.78 & 0.46 & 0.16 \\
R24 & 1.07 & 1.44 & 1.09 & 1.23 & 1.02 & 0.84 & 0.94 & 0.60 \\
M6 & 2.61 & \textbf{2.23} & 1.19 & 1.32 & 0.94 & 0.63 & 0.47 & 0.49 \\
M12 & 1.64 & 1.75 & 0.99 & 1.27 & 1.07 & 0.60 & 0.71 & 0.51 \\
M24 & 0.99 & 1.16 & 1.31 & 1.18 & 0.93 & 1.05 & 1.03 & 0.51 \\
Ensemble & 1.96 & 1.92 & \textbf{1.36} & \textbf{1.44} & \textbf{1.24} & 1.06 & 0.92 & 0.76 \\
\bottomrule
\end{tabular}
\caption{2025 Rank IC (\%).}\label{tab:appendix-2025}\end{table}
\clearpage
\begin{table}[H]\centering\small\begin{tabular}{lrrrrrrrr}
\toprule
Model & 5m & 10m & 15m & 20m & 25m & 30m & 60m & 120m \\
\midrule
C6 & 0.63 & \textbf{0.92} & 0.45 & 0.84 & 0.86 & 0.54 & 0.40 & 0.22 \\
C12 & 0.41 & 0.18 & 0.42 & 0.80 & 0.72 & 0.19 & 0.63 & 0.04 \\
C24 & 0.80 & 0.43 & 0.27 & 0.59 & 0.33 & 0.68 & 0.13 & 0.17 \\
U6 & 0.30 & 0.65 & 0.63 & 0.63 & 0.64 & 0.36 & 0.80 & 0.60 \\
U12 & 0.76 & 0.75 & 0.55 & 0.68 & 0.51 & 0.49 & 0.38 & 0.26 \\
U24 & \textbf{1.33} & 0.78 & 0.76 & 1.03 & 0.98 & 0.94 & \textbf{1.05} & 0.50 \\
R6 & 1.00 & 0.91 & 0.63 & 0.82 & 0.88 & 0.65 & 0.21 & -0.11 \\
R12 & 0.44 & 0.65 & 0.65 & 1.18 & 0.81 & 0.82 & 0.62 & 0.05 \\
R24 & 1.26 & 0.84 & 0.53 & 1.08 & 0.61 & 0.67 & 0.14 & 0.61 \\
M6 & 0.71 & 0.51 & 0.61 & 0.88 & 0.89 & 0.68 & 0.46 & 0.23 \\
M12 & 0.52 & 0.72 & 0.56 & \textbf{1.24} & 1.00 & 1.07 & 0.59 & 0.43 \\
M24 & 1.02 & 0.55 & 0.59 & 1.03 & \textbf{1.26} & \textbf{1.15} & 0.22 & \textbf{1.03} \\
Ensemble & 1.01 & 0.88 & \textbf{0.78} & 1.14 & 1.09 & 0.89 & 0.67 & 0.47 \\
\bottomrule
\end{tabular}
\caption{2026 Rank IC through 20 October (\%).}\label{tab:appendix-2026}\end{table}
\begin{table}[H]\centering\small\begin{tabular}{lrrrrrrrr}
\toprule
Model & 5m & 10m & 15m & 20m & 25m & 30m & 60m & 120m \\
\midrule
C6 & 3.33 & 2.02 & 0.91 & 1.26 & 1.13 & 1.38 & 1.03 & 1.46 \\
C12 & 3.26 & 2.34 & 1.61 & 1.37 & 1.67 & 1.59 & 0.99 & 0.58 \\
C24 & 3.34 & 1.99 & 1.31 & 1.34 & 1.34 & 1.28 & 0.99 & 0.85 \\
U6 & 3.18 & 2.10 & 1.44 & 1.68 & 1.27 & 1.17 & 0.54 & 0.99 \\
U12 & 3.55 & 1.88 & 1.11 & 1.41 & 1.53 & \textbf{1.68} & 1.34 & \textbf{1.58} \\
U24 & 3.11 & 1.49 & 0.45 & 1.26 & 1.19 & 1.55 & 1.19 & 1.00 \\
R6 & 3.63 & 2.16 & 1.22 & 1.43 & 1.25 & 1.31 & 0.85 & 0.86 \\
R12 & 3.36 & 2.11 & 1.29 & 1.43 & 1.55 & 1.07 & 0.70 & 0.31 \\
R24 & 3.19 & 2.26 & 1.56 & 1.79 & 1.55 & 1.36 & 1.31 & 1.06 \\
M6 & 3.50 & 2.15 & 0.59 & 1.15 & 1.12 & 1.03 & 0.80 & 0.65 \\
M12 & 3.07 & 2.10 & 1.40 & \textbf{1.80} & \textbf{1.86} & 1.00 & 0.98 & 0.51 \\
M24 & 3.36 & 2.18 & \textbf{1.67} & 1.69 & 1.54 & 1.65 & \textbf{1.53} & 1.10 \\
Ensemble & \textbf{3.72} & \textbf{2.43} & 1.49 & 1.72 & 1.66 & 1.65 & 1.32 & 1.20 \\
\bottomrule
\end{tabular}
\caption{H1 2025 Rank IC (\%).}\label{tab:appendix-2025h1}\end{table}
\clearpage
\begin{table}[H]\centering\small\begin{tabular}{lrrrrrrrr}
\toprule
Model & 5m & 10m & 15m & 20m & 25m & 30m & 60m & 120m \\
\midrule
C6 & 0.58 & 1.05 & 0.59 & 1.25 & 0.83 & 0.52 & 0.42 & 0.42 \\
C12 & 0.53 & 0.86 & 1.03 & 1.23 & 0.67 & 0.30 & 0.52 & -0.01 \\
C24 & -0.70 & 0.91 & 1.21 & 0.78 & 0.39 & 0.35 & 0.35 & 0.19 \\
U6 & 0.19 & 0.23 & 0.37 & 0.51 & 0.74 & 0.21 & 0.44 & 0.27 \\
U12 & -0.78 & 1.04 & 1.05 & 1.16 & \textbf{0.90} & \textbf{0.54} & \textbf{0.77} & \textbf{0.76} \\
U24 & 0.18 & 0.55 & 1.01 & 0.36 & 0.26 & 0.04 & 0.30 & 0.27 \\
R6 & 1.68 & 2.04 & 1.38 & 1.18 & 0.81 & 0.33 & 0.24 & 0.49 \\
R12 & 0.28 & 1.41 & 0.68 & 1.02 & 0.89 & 0.50 & 0.22 & 0.01 \\
R24 & -0.96 & 0.65 & 0.65 & 0.69 & 0.51 & 0.34 & 0.59 & 0.16 \\
M6 & \textbf{1.76} & \textbf{2.30} & \textbf{1.76} & \textbf{1.48} & 0.76 & 0.25 & 0.16 & 0.33 \\
M12 & 0.28 & 1.42 & 0.60 & 0.76 & 0.32 & 0.23 & 0.45 & 0.52 \\
M24 & -1.27 & 0.18 & 0.98 & 0.70 & 0.36 & 0.47 & 0.54 & -0.04 \\
Ensemble & 0.28 & 1.44 & 1.23 & 1.18 & 0.84 & 0.50 & 0.54 & 0.35 \\
\bottomrule
\end{tabular}
\caption{H2 2025 Rank IC (\%).}\label{tab:appendix-2025h2}\end{table}

\clearpage\section{Secondary diagnostic: naive target turnover}
Trading-performance analysis is not the focus of this study. A forecast-only target exercise illustrates why the naive portfolios would be poor evidence of attainable performance. At each 30-minute origin, the equal ensemble's 30-minute prediction forms three gross-two targets: equal weights within positive/negative forecasts; weights proportional to forecast magnitude separately on each side; and equal-weight top/bottom quintiles. They are reset to flat overnight. Target one-way turnover is
\begin{equation}
TO^{\mathrm{one}}_d=\frac12\sum_{t\in d}\sum_i|w^{\mathrm{target}}_{i,t}-w^{\mathrm{previous\ target}}_{i,t}|,
\end{equation}
including opening deployment and closing liquidation. Two-way turnover is twice this quantity. Weights are fractions of NAV; these are not fractions of the gross book. The calculation ignores price drift, partial fills, capacity, costs and optimizer constraints.
\begin{table}[H]\centering\caption{Mean daily 30-minute target turnover of the equal ensemble.}\begin{tabular}{lrr}
\toprule
Portfolio & One-way / NAV (\%) & Two-way / NAV (\%) \\
\midrule
Equal-sign & 1,042.9 & 2,085.7 \\
Prediction-weighted & 1,332.6 & 2,665.3 \\
Q5--Q1 & 1,466.7 & 2,933.5 \\
\bottomrule
\end{tabular}
\end{table}
Daily two-way target turnover exceeds 2,000\% of NAV. That motivates cost-aware execution and slower target adjustment, but is not itself a realized trading statistic. The calculation does not provide average holding periods: opening and closing both contribute to traded volume, and partial rebalances can occur without closing a position.

No cumulative return, net profit or Sharpe ratio is reported for these targets. High turnover and incomplete held outcomes make those naive quantities unsuitable as the main evidence. Any later economic assessment should use a separately specified execution and price-marking policy.

\end{document}